\documentclass[acmsmall]{acmart}

\setcopyright{rightsretained}
\acmJournal{PACMHCI}
\acmYear{2024} \acmVolume{8} \acmNumber{CSCW1} \acmArticle{167} \acmMonth{4} \acmPrice{}\acmDOI{10.1145/3641006}

\usepackage{makecell}
\usepackage{longtable}

\received{July 2023}
\received[revised]{October 2023}
\received[accepted]{November 2023}

\begin{document}

\title{Form-From: A Design Space of Social Media Systems}


\author{Amy X. Zhang}
\affiliation{%
  \institution{University of Washington}
  \city{Seattle}
  \state{WA}
  \country{USA}
}
\email{axz@cs.uw.edu}
\orcid{0000-0001-9462-9835}

\author{Michael S. Bernstein}
\affiliation{%
  \institution{Stanford University}
  \city{Stanford}
  \state{CA}
  \country{USA}
}
\email{msb@cs.stanford.edu}
\orcid{0000-0001-8020-9434}

\author{David R. Karger}
\affiliation{%
  \institution{Massachusetts Institute of Technology}
  \city{Cambridge}
  \state{MA}
  \country{USA}
}
\email{karger@mit.edu}
\orcid{0000-0002-0024-5847}

\author{Mark S. Ackerman}
\affiliation{%
  \institution{University of Michigan}
  \city{Ann Arbor}
  \state{MI}
  \country{USA}
}
\email{ackerm@umich.edu}
\orcid{0000-0001-9727-1664}

\renewcommand{\shortauthors}{Amy X. Zhang, Michael S. Bernstein, David R. Karger, \& Mark S. Ackerman}

\begin{abstract}
 Social media systems are as varied as they are pervasive.  They have been almost universally adopted for a broad range of purposes including work, entertainment, activism, and decision making.   As a result, they have also diversified, with many distinct designs differing in content type, organization, delivery mechanism, access control, and many other dimensions. In this work, we aim to characterize and then distill a concise design space of social media systems that can help us understand similarities and differences, recognize potential consequences of design choices, and identify spaces for innovation. Our model, which we call \textit{Form-From}, characterizes social media based on (1)~the form of the content, either threaded or flat, and (2)~from where or from whom one might receive content, ranging from spaces to networks to the commons. We derive Form-From inductively from a larger set of 62 dimensions organized into 10 categories. To demonstrate the utility of our model, we trace the history of social media systems as they traverse the Form-From space over time, and we identify common design patterns within cells of the model.
\end{abstract}




\maketitle

\section{Introduction}

New social media platforms rise all the time. Even in the time we were finishing this paper, several platforms arose as competitors to the massive platform Twitter:\footnote{Twitter was also recently renamed to X. We refer to it as Twitter to reflect the time period when the research was conducted.} Mastodon pushed a decentralized approach, Twitter's co-founder announced Bluesky with a promise of per-user personalized moderation, Meta launched a competitor called Threads to leverage their user base, Substack pushed out a platform called Notes, and still others promise that their designs or moderation strategies will produce friendlier and more welcoming communities.

What makes these social media designs different from, or similar to, one another? Considering two popular platforms in 2024, the main Instagram feed is different from TikTok's For You feed---but how? Certainly Instagram is focused on images, whereas TikTok is focused on video, but we could swap Instagram's photos for TikTok's videos and the platforms would still feel and act differently. 

Given the many iterations and continually changing designs of social media systems, a comprehensive design space and succinct models for that design space are essential for mapping systems to articulate similarities and differences, charting the evolution in system designs, and ideating possible future systems. Through its evolution, CSCW has benefited from design space models such as the time-space matrix widely attributed to Johansen~\cite{johansen1988groupware} and Grudin~\cite{grudin1994computer}. Yet as social media has rapidly evolved in the last thirty years, today nearly all social media systems fall in the exact same ``different time-different place'' cell in the time-space matrix. So, social media researchers and designers are overdue for an update to our design models that better captures the similarities and differences in social media design.

In this paper, we pursue a model that supports understanding and design of social media. Our goal is to describe a design space that separates meaningful deep differences in social media design, while placing together designs that may appear separate at first glance but share deeper DNA than might otherwise appear. We place our focus on social media designs, which we define as systems that directly facilitate the sharing of user-generated content between users. Taking a wide lens on this definition, we include traditional social media such as Facebook, Twitter, and Reddit, but also other user-generated content platforms for communication such as group chat (e.g., WhatsApp), email, video discussions (e.g., Zoom), and semi-synchronous chat (e.g., Slack). In these systems, users engage with others primarily by adding their content to a growing repository of content in response to others’ content, forming cross-user dialogue.
This definition leaves out systems where users engage by successively iterating over the same content and overwriting each other’s work, as in collaborative document editing (e.g., Google Docs). We choose to focus  on differences in features and design because they have influence on specific classes of behaviors~\cite{kraut2012building} and are directly under the control of the platform designer, though of course they do not determine socio-technical outcomes.

\begin{figure}
    \centering
    \includegraphics[width=1.0\textwidth]{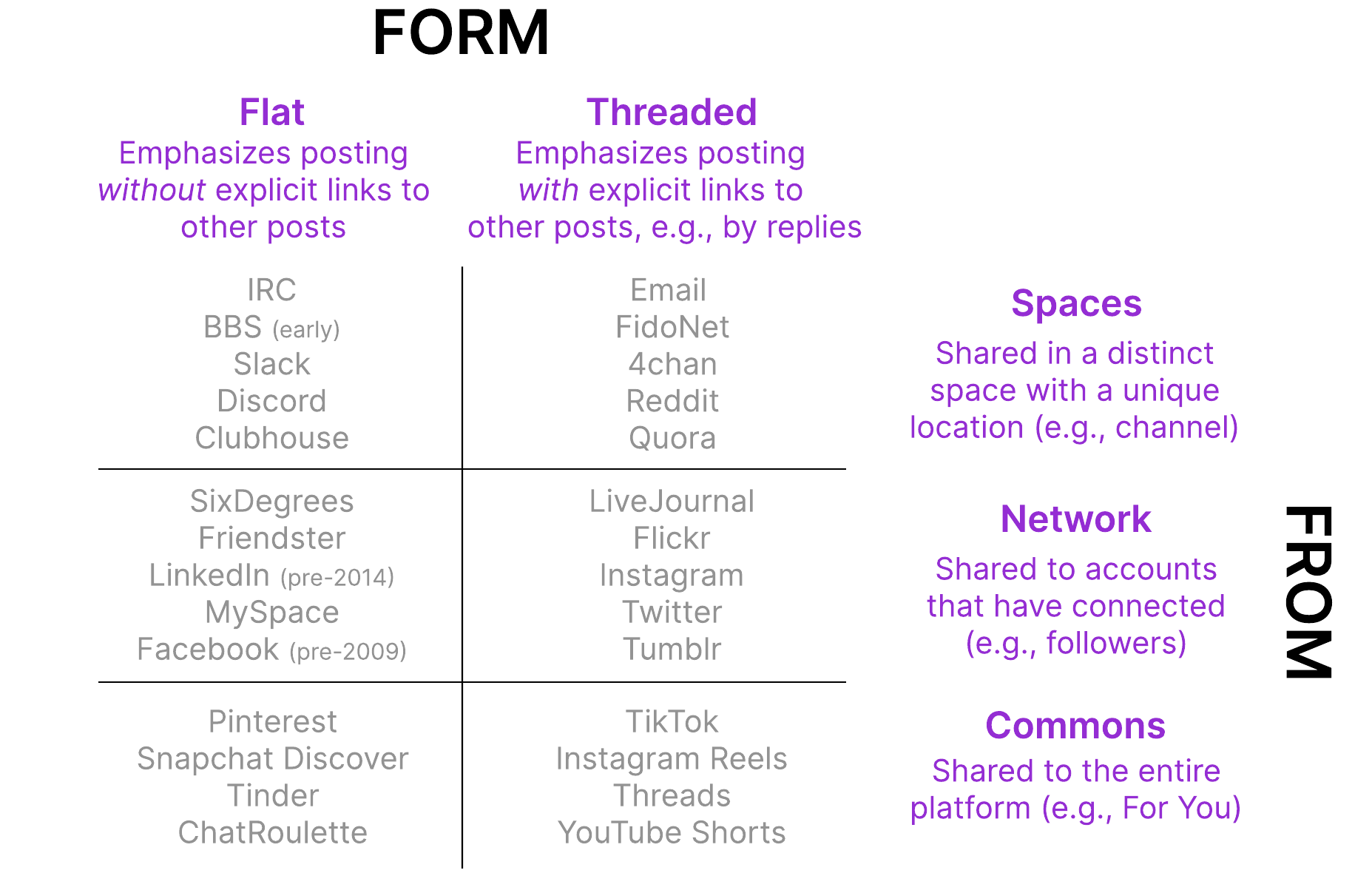}
    \caption{The Form-From model describes social media designs along two main dimensions: (1)~The \textit{form} or shape that the content takes, either flat as in a chatroom or threaded as in individual posts with linked comments, and (2)~\textit{from} where a user receives content, as from spaces such as channels, from networks such as a friend graph, or from a platform-wide commons such as an algorithmic For You page. For simplicity, examples in this figure are coded as of their design at their initial launch.}
    \label{fig:formfrom}
\end{figure}

We describe a model of social media designs that we call \textit{Form-From} (Figure~\ref{fig:formfrom}). We derived this model through an inductive method and purposive sampling of public platforms and social computing research systems. The Form-From model focuses on two main dimensions of difference. The first, the \textit{Form} dimension, describes the principal form (shape) of the main content on the platform: is the design \textit{threaded} 
by emphasizing explicitly linking posts to each other, for example through replies, or is it \textit{flat} where the design 
emphasizes posts that are not explicitly linked to each other, instead focusing interaction with the overall stream? The second, the \textit{From} dimension, describes the principal process by which content is delivered from one user to another: is content delivered to the user from spaces (e.g., channels, groups, or rooms) where all members in the space receive it, from (social) networks where each user independently curates a set of friend or follow edges that determines what they receive, or as a sample from the ``commons'' of all content on the platform? While platforms can blend elements of different options, they are often principally designed around one of the six cells in this 2x3 Form-From model. This model helps make comparative sense of platforms, for example placing Instagram's delivery as principally based on a social network, where TikTok's is principally sampling from the commons, despite both primarily delivering visual content. Likewise, we characterize Discord channels as being principally flat, vs. traditional forums as being principally threaded, despite both being text-based group discussion platforms that support threaded replies and comments on posts. In fact, when Discord launched a feature called ``forum channels'' that forced all comments to be in reply threads, it flipped that bit from flat to threaded, deeply changing the social dynamics in those channels.

We derived Form-From by first generating a larger set of 62 dimensions that aims to summarize the key types of differences between social media systems via an inductive process covering a large number of industry and research systems. We hierarchically grouped these 62 dimensions into ten categories:
Discussion Structuring (simplified to ``Form''), Delivery (simplified to ``From''), Temporality, Audience, Membership, Moderation, Governance, Extensibility, Signaling, and Content. While the Form-From model captures what we argue are two key dimensions of variation across social media designs, these ten categories and 62 dimensions serve as a fuller treatment for comparing similar designs. For example, while Twitter, Mastodon, and Bluesky all occupy the threaded-network cell of the Form-From model, they differ in their approaches to Governance, Moderation, and Extensibility.

We also apply our Form-From model to make sense of the history and possible futures of social media design. First, we trace popular social media platforms as they arise over time and observe a pattern of migration from early ``flatspace'' systems in the 1970s--1980s, to threaded spaces, to networks---again first flat, and then later threaded---in the 1990s--2010s, and now toward threaded commons designs. We also explore design patterns that occur within cells of the model, for example the prevalence of a highly visible channel list navigation pane in the flat space cell.

Our contributions in this paper are:
\begin{enumerate}
    \item The Form-From model, which enables comparative analysis of designs and design patterns in social media;
    \item A finer-grained description of the design space using ten categories encompassing 62 dimensions for higher-resolution comparisons;
    \item Applications of these models to analyzing social computing and social media.
\end{enumerate}
To begin, we will review the history of design space models in HCI/CSCW and design more generally. We will then introduce our method  and our models of the social media design space beginning from the coarsest (Form-From) to finest (the raw 62 dimensions).  Finally we walk through some applications of the models.

\section{Background}

\subsection{Design Spaces}
Models for technical systems have a long history in the HCI/CSCW literature.  As simplifications of the characteristics of HCI systems or of HCI uses, models have proved exceedingly helpful.  
Roughly, there are two strands of work in HCI/CSCW that develop and analyze design spaces and models for those design spaces.  The most detailed ones are based on Simon’s work~\cite{simon1996sciences}, where design was viewed as search through a problem space. 
Researchers following the Simonian view have largely construed design spaces as  parameterized spaces of characteristics that support algorithmic search.  (While design spaces as a concept were also considered in engineering and operations research, e.g., Zwicky~\cite{zwicky1967morphological} and Jones~\cite{jones1992design}, Simon’s descendants were the ones who brought design space exploration to HCI/CSCW.)  In this view, a complete design space is literally the space of all possible designs, where each dimension of the space is a technical feature or characteristic for some collection of systems.

The earliest such work in HCI/CSCW that we know of is Card, Mackinlay, and Robertson’s work on input devices~\cite{card1990design,card1991morphological}.  There~\cite[p. 101]{card1991morphological}, the authors argue they found it necessary to ``represent the designs as points in this design space [of input devices], some parametric representation [must be] determined that can represent the central idea of particular designs.''  Similar work has continued in HCI/CSCW, largely in the area of interaction devices and visualizations.  For example, some researchers laid out the design space of windshield display apps for automobiles~\cite{10.1145/2858036.2858336} or conversational in-vehicle information systems~\cite{Braun2017ADS}. Other work laying out design spaces within HCI/CSCW include head-mounted displays~\cite{Hirzle2019ADS} and electronic health record visualizations~\cite{Belkacem2017ExploringAD}. 

Outside of HCI/CSCW, design space analyses of hardware or devices flourish, tending to consider well-contained sets of characteristics.  Example analyses include multimodality as a systems consideration~\cite{nigay1993design}, self-adaptive systems~\cite{Brun2010ADS}, mobile medical devices~\cite{Kulkarni2007RequirementsAD}, and communication architectures for systems-on-a-chip~\cite{Lahiri2000EfficientEO}.  In practice, the feature sets are relatively small and therefore tractable to this approach. 

The second strand of design space work is based more on Schön’s view of design~\cite{schon1984reflective}. A Schönian view  emphasizes the serendipity and messiness in the design process.  In this view, a design space is, in Gaver's~\cite{gaver2011making} terms, the curated collection of design concepts. ``Design space'' as a concept is different than in the Simonian view: ``The ideas presented in these artifacts are said to occupy a metaphorical landscape of design opportunities, rather than a Cartesian space of possible designs; and their purpose is to inspire rather than prescribe design choices''~\cite{dove2016argument}.  Exploration, then, is through a set of design prototypes  that loosely define a space of potential designs. 
In practice, what people mean when they talk about this type of design space analysis is often the use of a highly simplified, conceptual, and emergent model through which they can explore design alternatives and potentially their rationales (e.g.,~\cite{10.1145/1375761.1375762}).   While this use of ``design space'' points to an important consideration in HCI/CSCW, especially in Research through Design~\cite{Zimmerman2007ResearchTD}, in this paper we take a Simonian view and focus on demarking the technical features.

We note that some Simonian analyses also focus on a limited number of key dimensions.  For example, Feng, Yao, and Sadeh’s~\cite{feng2021design} purpose is to ``help researchers and practitioners better understand the key dimensions to be considered when designing privacy choices for their systems.''  They therefore focus on what they consider to be the most important five dimensions in their design space.  Simplifying a complex design space is likely useful for those designing systems; we will discuss this further below.

We also recognize the considerable amount of work in HCI/CSCW that has focused on socio-technical design spaces that include elements of user tasks or requirements as well as technical features.  As one example among many, Lee and Paine~\cite{lee2015matrix} contribute a model of all coordinated distributed work.  Their model includes technical capabilities, but it also includes the planned permanence of distributed action and the turnover rate of participants. While we recognize the importance of this and similar work, again the emphasis in this paper is on technical features.

In summary, the literature presenting models of design spaces in HCI/CSCW has been split.\footnote{Some researchers have attempted to reconcile the two, pointing out that Simon also wrote about the ``search'' in a ``design space'' in a metaphorical manner (e.g., in Sciences of the Artificial~\cite{simon1996sciences}).  We also note that Simon wrote at the same time as Zwicky~\cite{zwicky1967morphological}, with his morphological analysis, which is closer to what we have labeled as Schönian design spaces.  Nonetheless, we find it useful analytically  to separate the Simonian and the Schonian as to highlight the differences.}  Given we want to be able to relate different social media designs conceptually against one another, along with having the additional goal of finding new types of social media systems, we believe that laying out a detailed description of the technical characteristics in a Simonian design space analysis will be the more valuable approach here.

\subsection{Design Spaces for Social Media Systems}

\begin{figure}
    \centering
    \includegraphics[width=1.0\textwidth]{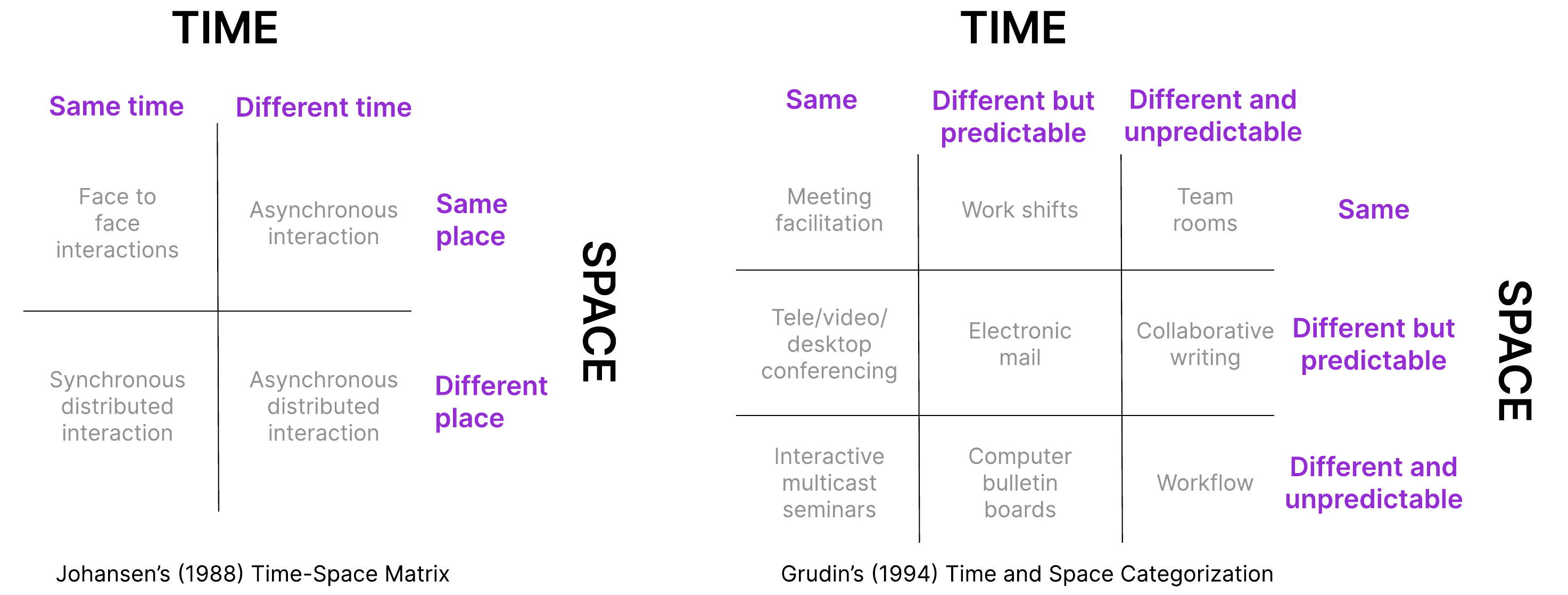}
    \caption{Johansen's time-space matrix~\cite{johansen1988groupware} and Grudin's time-space categorization~\cite{grudin1994computer} are two classic models of CSCW systems. They both split the design space by whether the people are interacting synchronously or asynchronously, and by whether people are in the same physical location. Most modern social media systems fall into the same ``different time--different place'' cell of these models.}
    \label{fig:grudin}
\end{figure}

There is scant work examining the design space for social media systems.  Perhaps the most well-known model in CSCW, seen in Figure~\ref{fig:grudin}, is what is commonly known as Johansen’s time-space matrix~\cite{johansen1988groupware} for collaborative systems, consisting of a 2x2 matrix of location (same place-different place) 
versus time (same time-different time).\footnote{An even earlier version of this matrix can be traced back to DeSanctis and Gallupe~\cite{desanctis1984group} on Group Decision Support Systems (GDSS). They had a 2x2 matrix of ``dispersion of group members'' (close proximity or dispersed) versus ``duration of decision-making session'' (limited or ongoing). This matrix went on to inspire both Johansen and Grudin and became popularized as Johansen's time-space matrix.}  Grudin later expanded upon this to be a 3x3 matrix~\cite{grudin1994computer}. 
The Johansen and Grudin time-space matrices could be considered types of ``design space'' analyses, albeit relatively simple ones. At the time, these models were foundational in guiding both research and commercial products by clarifying the types of systems that might be constructed or considered.   

However, we believe they 
are not well suited to discussing social media systems, which
now encompass a substantially  larger range and complexity. When we consider widely-used systems that facilitate social communication today, some are commonly used for work within organizations
(i.e., Zoom, Slack, email), which was the focus of early CSCW research.  But others, used just as widely if not more, primarily host activities outside of traditional groupwork 
(i.e., Facebook, Instagram, TikTok, WhatsApp). When we do consider this broader set of systems, a wide range of the most popular social communication tools---Slack, email, Facebook, Instagram, TikTok, and WhatsApp---which have significant diversity of function, would all fall into the same cell of Johansen’s matrix: remote and asynchronous. 

There are relatively few other design space analyses of social media systems.  A recent development analyzing design spaces for social media systems is Rajendra-Nicolucci and Zuckerman~\cite{rajendra2021illustrated}.  Rajendra-Nicolucci and Zuckerman present many abstracted ``logics'' that can inform the design of social media systems.  Each of their illuminating analyses, which they argue create something similar to a bestiary of types, is informed by concrete examples.  However, there is not a clear distinction of dimensions of analysis, as one might want in a Simonian design space analysis. 
Other analyses focus on examining specific dimensions of the design space.  These include  Smith et al.~\cite{smith2000conversation}  as well as Gilbert~\cite{gilbert2012designing} among others.  Gilbert characterizes social media systems through the prism of different kinds of awareness, demonstrating how different designs about social translucence derive from different decisions about who can see what between triads of users. Smith, Cadiz, and Burkhalter~\cite{smith2000conversation} examined new types of conversational organization, such as what they called threaded chat and conversation trees.  These are all valuable additions to understanding the design space of social media systems, but we believe it would be valuable to see a more complete representation of the entire design space.

It is time to develop a model better-matched to systems that incorporate social communication; we need a new Simonian analysis and design space that adequately encompasses and distinguishes the major systems of today.

\section{Method}

Our process of generating a design space was inductive, beginning with a survey of relevant research and industry systems, continuing with the generation of a large number of low-level dimensions of variation between those systems, then refining iteratively to reduce those dimensions to a small number of central concerns for the design space, and finally holding up the design space against the actual systems to test its validity and expressivity.

\subsection{Scope and Definitions}

To make our modeling effort tractable, we limit our scope to social media systems: systems that facilitate the sharing of user-generated expressive content, or media, from one user to others.\footnote{Merriam Webster definition of social media: forms of electronic communication (such as websites for social networking and microblogging) through which users create online communities to share information, ideas, personal messages, and other content (such as videos).} In such systems, social activity revolves around users creating content \textit{ab initio} or in response to other users’ content, primarily in an appending fashion. Such social media systems include systems that would not fit neatly into classic CSCW/groupware, including systems that are designed to be used primarily for entertainment (i.e., by surfacing highly engaging content). Conversely, some CSCW systems, such as Microsoft Excel with its multi-user cell editing capabilities, or Wikipedia article pages with their collaborative editing of documents, would not fit into our concept of a social media system due to the attention to editing over appending and the lack or paucity of feature support for user responses or conversations.  Our scope covers an enormous number of CSCW/social media systems, but is sufficiently contained to provide analytical traction.

Consistent with this goal, our analysis focuses on each system’s designs for sharing, publishing, and communicating content, and gives less attention to the (possibly sophisticated) affordances for creating the content to be shared.

Even beyond social media, many CSCW systems do enable some form of back-and-forth messages in a group or broadcasting of content to a large audience, since having a communication channel is typically necessary to support coordination. But importantly, we distinguish the communication portion of these systems from their central arena for coordination (more detail to come in Section~\ref{sec:surfaces})---so while Wikipedia Talk pages would constitute a social media \emph{surface}, Wikipedia articles would not. 

\subsubsection{Surfaces}\label{sec:surfaces}

Today's social media platforms are multifaceted and incorporate multiple sets of system designs for social communication. Due to these complexities, we also classify distinct ``surfaces'' within a system instead of categorizing systems as a whole. Many platforms today incorporate a combination of different internal subsystems, each with different user-facing system designs, with more or less integration; for instance, ``Facebook'' as a whole mashes together Facebook Groups, Messenger, Pages, and the original Facebook News Feed. It would be difficult to classify Facebook as a whole given the differences in how each of the different internal systems work, even though they can all be accessed from the same website. 

Within a product surface, there is a common way that social communication happens and a common vocabulary for describing social actions and user-facing objects that is shared among all users.
When two product surfaces appear in the same social media platform, there may be no integration outside of a common system for user identities and accounts that lets a user easily move from one product surface to another. Or there may be significant integration, such that communications that happen in one product surface get transmitted to or accessed via another. 

In our analysis, we spent substantial time trying to determine when systems had sufficiently changed to be considered separate surfaces.  This problem is important only when considering how systems change over time (e.g., gathering additional capabilities so as to compete with novel systems) and in mapping dimensions to systems.  Otherwise, determining the exact instance of a long-standing system (e.g., Facebook after likes and before likes) is not crucial to the development of our design space; as long as we have considered the various instances of a system, we have obtained all of the attributes across the instances to help make our design space more comprehensive.

Overall, in this paper, we have tried to be precise when we mention a system to name the specific product surface we are analyzing; however, in cases where we perceive the system to have a primary surface, we use the system name as a shorthand for the surface. For instance, when we use the term ``Instagram,'' we are referring to the Instagram Feed; when talking about secondary Instagram surfaces, we use more precise names such as ``Instagram Stories'' or ``Instagram Reels.''

\subsubsection{Technical System Design}
In addition, we choose to focus our taxonomization of product surfaces according to their apparently  intended usage
rather than the way they may have been appropriated by their users for different aims. For instance, while users may appropriate collaborative editing documents for back-and-forth discussion by taking turns editing text into one location, they are not using built-in technical features that were designed to facilitate communication.

\subsubsection{Principal Uses}
Finally, social platforms are complex, oftentimes involving multiple ways for users to interact with each other, even when focusing on a specific product surface. For instance, on Slack, a user can interact with others by reacting with emojis, typing a new message, starting a thread, or several other ways. Is Slack a flat system with some threaded elements, or a threaded system with some flat elements? In these cases, we try to determine the central or principal way that users interact with each other within the system in terms of the focus of the system design, and then label according to that principal use, while ignoring the ways that are more peripheral. In the case of Slack, we chose the social communication that happens in the text chat. Likewise, while Slack has added support for threads, its principal use in most scenarios remains a flat stream of messages---except in extremely high membership channels, most discussion occurs outside of threads. We acknowledge that these decisions are up for debate, and we note where we make these decisions. In addition, we are not claiming that these peripheral modes of communication are unimportant. But by focusing on the dominant functionality we can more easily classify existing systems into higher level patterns.

\subsection{Inductive Process}

\begin{figure}[tb]
    \centering
    \includegraphics[width=1.0\textwidth]{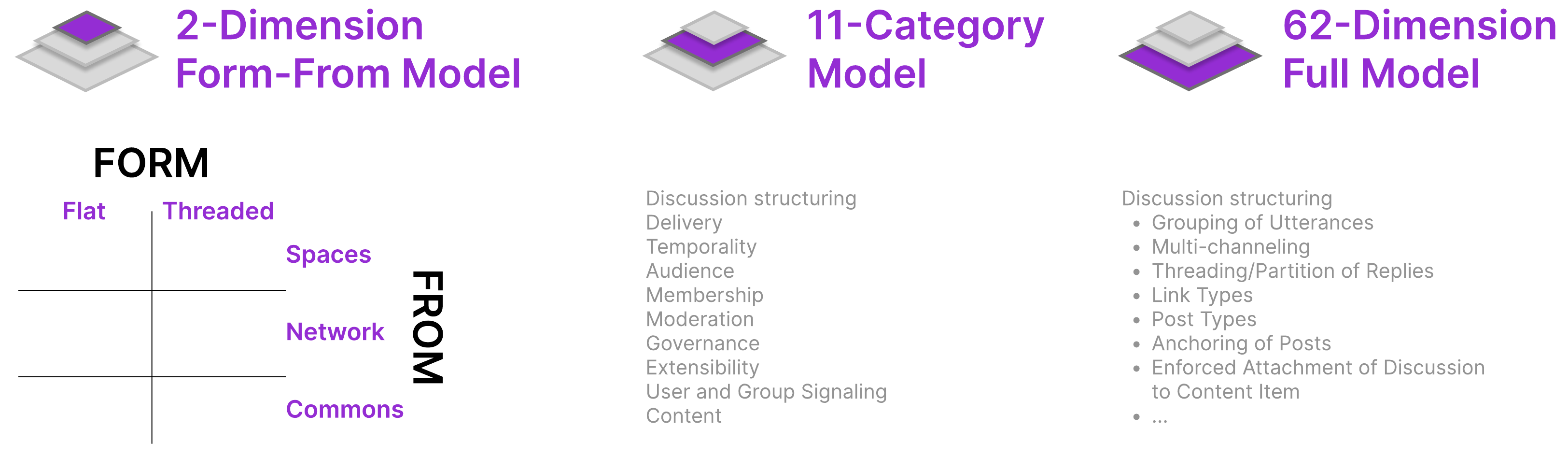}
    \caption{The Form-From model (left), the focus of our paper, is a distillation of a larger design space that we articulated and synthesized through our inductive process. We began with a review of systems and platforms to produce a large, maximal model that articulates 62 dimensions that we considered salient (right). We then iteratively grouped those dimensions hierarchically into 11 categories (middle), and finally selected and distilled them into two dimensions for Form-From.}
    \label{fig:formfrom}
\end{figure}

Our first step was to gather contemporary and historical social media systems that our design space should describe. We sought to include a wide range of such systems, including systems that are in industry, community-owned, and research. Rather than a systematic literature review, we performed purposive sampling~\cite{palinkas2015purposeful}, a standard interpretivist technique in which we selected research papers and public platforms that we considered were particularly influential, or had articulated interesting new areas of the social media design space.
Papers were drawn from the CSCW and social computing literature, e.g., the CHI and CSCW conferences. 

Second, we developed a large number of \textit{design dimensions} that described differences between the systems. In generating these dimensions, we focused on differences in the design or technical affordances of the systems. Each dimension was categorical, requiring zero or more values. Example dimensions included: are replies broadcast to the entire group, or not? What powers do administrators have? Are comments required to be anchored to a piece of content (e.g., photo or video), or not? In creating these dimensions, we sought ``differences that make a difference'' in the system design. Our focus was on technical and design system decisions, rather than sociotechnical outcomes: in other words, axes could capture what user-visible features varied across platforms, but not how users of those systems organized themselves using those features.

We inductively grouped these design dimensions into higher level \textit{categories}, for example Audience, Membership, and Moderation. Following a process similar to thematic analysis, we coded dimensions with potential higher level categories, grouped them, reflected on the categories, and regrouped. The goal of this process was to provide a higher level organizational structure to the dimensions. 

We then took these dimensions and categories and used them to identify additional systems that were not included in our initial set. To do so, we sampled underpopulated areas in the multi-dimensional space, then sought out systems that might fill those holes. 
For instance, the consideration of structured argumentation systems in both research (e.g., ArguLens~\cite{Wang2020ArguLensAO}, MIT Deliberatorium~\cite{Klein2012HowTH}, ConsiderIt~\cite{Kriplean2012SupportingRP}) and industry (e.g., Kialo,\footnote{\url{https://www.kialo.com}} Arguman,\footnote{\url{https://github.com/arguman/arguman.org}} Pol.is\footnote{\url{https://pol.is}}) led us to incorporate dimensions that capture when systems have a constrained set of post types and link types. 
In addition, we reflected on systems that did not align with any of the currently existing options for a given dimension and introduced additional options to cover them.

We continued this iterative process until theoretical saturation. In our iterative process, the additional systems prompted revision of the dimensions, which prompted revision of the higher-level categories, which then led us to continue to generate additional systems. In total over a period of around six months, the authors individually annotated and discussed over 100 systems in total, with eleven
of them receiving full annotations across all 62 dimensions: WhatsApp, 4chan, Quora, Zoom, Slack, OhYay, email to specific individuals, mailing lists, Reddit, Twitter, and the New York Times comments section.

\subsection{External Validation of Dimensions Against Systems}

As the dimensions and categories stabilized, we recruited four undergraduate students at our institutions to apply the full 62-dimensional design space coding to a series of 15 concrete systems with which they all had some familiarity:   Facebook Messenger, Twitter, GroupMe, Discord, EdStem, Piazza, Google Hangouts,  Google Docs' inline commenting, Clubhouse, Instagram, Instagram Stories, YikYak, Reddit, Snapchat, and TikTok. 
Initially, all four students coded one system surface very familar to all---Instagram Stories.  They then met as a group with two of the researchers to discuss points of confusion and disagreement in the codebook. After gaining a better understanding of the dimensions through this exercise, each student went on to separately and individually code one system at a time, focusing on systems with which they were familiar and that seemed different from what had already been coded. We would then meet as a group to discuss the codes. As these students identified unclear decision points, or systems that did not fit the existing set of dimensions, we further refined the dimensions and categories. We stopped after the students collectively coded 15 systems, as the codebook had stabilized at that point.  We decided that we had achieved theoretical saturation when we saw no new dimensions or values.  This is analogous to theoretical closure in interpretivism; no additional valuable data suggests that data collection can end.

Our final full set of dimensions included 62 dimensions along ten categories. A list of the full dimensions and values
appears in Appendix~\ref{adx:model}.

Noting that 62 dimensions, and even ten categories, is a lot to remember, and that simpler models can be more analytically tractable and powerful, we 
then (again iteratively) distilled the ten-category model to two dimensions---From and Form---that we felt captured the most salient dimensions of technical design. More so than any of the other dimensions,
these were the two we turned to again and again for discussion and that we noticed having an impact in multiple places across our 62-dimension space. Finally, through discussion, we arrived at the 2--3 options within each dimension that best divided the systems into meaningful and coherent groupings.

\section{The Form-From Model}

\begin{figure}
    \centering
    \includegraphics[width=1.0\textwidth]{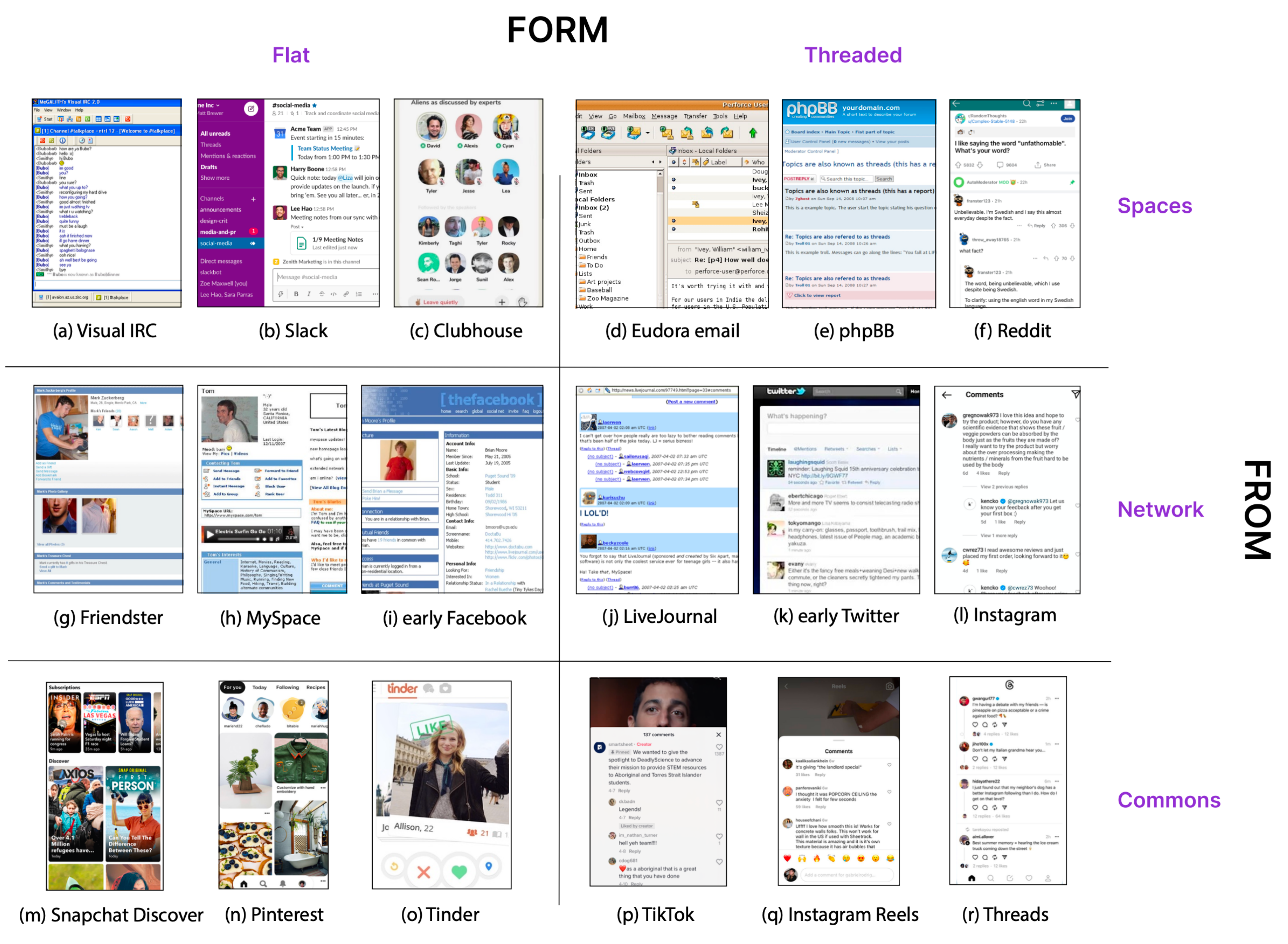}
    \caption{Examples of systems both old and current in each cell of the Form-From model.}
    \label{fig:examples}
\end{figure}

\subsection{2x3 Model}

Our highest level model of content sharing systems distinguishes between two major dimensions: \emph{Form} categorizes on the shape or structure of the principal unit of content, and \emph{From}, on how content is delivered to users. Example screenshots of different systems in each cell of the 2x3 can be found in Figure~\ref{fig:examples}.

\subsubsection{Dimension 1: The Form of Content}

Our first dimension describes the way that social content 
is organized, or what form the content takes at a high level.  The two  Forms are \emph{threaded} and \emph{flat}.
In a threaded system, the design emphasizes explicitly linking posts to each other, for example through replies.
Example threaded designs include:
\begin{itemize}
    \item Twitter, Mastodon, Threads, and Bluesky, which center attentions, replies, and reshares around individual posts, and where replies are posts that can be reshared much like top-level posts;
    \item YouTube, SoundCloud, TikTok, and other video or audio sharing sites, which privilege individual pieces of media and collect comments and reactions on them;
    \item Reddit, which orients its design around upvotes and replies to top-level posts;
    \item Traditional forums, which organize top-level posts chronologically and allow replies to them.
\end{itemize}
In contrast, in a \textit{flat} design, the design 
emphasizes posts that are not explicitly linked to each other, instead focusing interaction on the overall stream. Example flat designs include:
\begin{itemize}
    \item Traditional chatrooms such as IRC, which feature an undifferentiated stream of chats;
    \item Instant messaging platforms such as AOL Instant Messenger, Yahoo\! Messenger, WhatsApp, and Facebook Messenger, which similarly focus attention on the overall stream of chats rather than individual posts;
    \item Initial designs of platforms such as Facebook, MySpace, and LinkedIn, which showcased a stream of comments on a user's wall rather than enabling threaded replies;
    \item Video and audio livestreams such as Zoom, Twitch, YouTube Live, and Twitter Spaces, which are uncut and ongoing streams of video and audio;
    \item Instagram Stories, Snapchat, and other ``stories'' systems where users view a stream of video clips or images that disappear after a short period of time.
\end{itemize}

In some ways, the Form dimension correlates with the synchronous-asynchronous divide in the ``time'' dimension of the Johansen matrix, in that systems where people predominantly interact synchronously tend to appear in the Flat category, while systems where people predominantly interact asynchronously tend to appear in the Threaded category.
However, we argue that in today’s social media, time has become a less useful way to distinguish systems, as a vast number of systems are used today in a semi-synchronous fashion, making the synchronous-asynchronous divide more of a continuum. For instance, instant messaging, chat systems, and multi-user editors all appear in the synchronous side of the Johansen matrix~\cite{johansen1988groupware}---yet all of these systems can be used asynchronously and in many cases are  predominantly used asynchronously today~\cite{Wang2017WhyUD}. 
Instead of focusing on usage, which can change as norms change, our focus on the fundamental shape of user-to-user interaction, as conveyed by the underlying design and data model, enables us to more cleanly separate systems and also to pinpoint what design choices facilitate certain kinds of interactions. For example, while both chat and email can be and are used both synchronously and asynchronously, they seem optimal for opposite modalities.  Chat works well in a synchronous setting: since each post is typically replying (if at all) to the one immediately previous, there is little need for presenting complex reply threads and a flat chronological stream offers a simpler user experience.  In contrast, email works well for asynchronous conversations, because the explicitly represented threading structure supports navigation of multiple distinct conversations overlapping in 
time.\footnote{Temporality does emerge as an orthogonal category in our 10-category expanded model, with many facets that go beyond synchronous-asynchronous interaction.}  Conversely, if any one channel is going to be used for multiple simultaneous conversations, some form of threading is generally required in order to keep those conversations from jumbling together.  

We now describe each option in more detail and mention common subspecies within each.

\paragraph{Threaded Systems}
In this option, the design emphasizes explicitly linking posts to each other, for example through replies.
Each piece of content can be a post sent by a user that kicks off a series of posts that are all linked back (possibly indirectly) to the initial post.
Meta's system Threads, appropriately, is threaded by this definition.

There are different subspecies within this distinction. 
For example, threaded designs often have a form that we call \emph{headed}, as top-level content is treated differently from replies. Examples include Reddit and Facebook posts, where the top-level posts are quite distinct from replies; for instance, only top-level posts can appear ranked in one's feed.
The ``head'' that kicks off discussion can also be its own distinct form of content, such as a news article, a document, or a video, that receives comments on it. Examples include YouTube and TikTok, where the initial content is a video, Instagram, where the initial content is an image, or comments sections of news organizations such as the New York Times, where the initial content is a news article.  In general for these platforms, the only way to reach the comments hanging off a particular head is via that head content itself---e.g., the comments for a YouTube video can only be found on that video's page.
For these kinds of platforms, a further distinction involves whether there are one or more sub-threads that branch from individual replies. For instance, Reddit and Twitter allow for infinite branching of replies into sub-threads, while other platforms only allow a single thread or one or two levels of branching.\footnote{This distinction surfaces as two of the possible options in the dimension ``Threading/Partition of Replies'' in our 62-dimension model (\ref{dim:reply-partition}).} 

In contrast, in \emph{headless} designs, responses to posts are first-class objects in their own right, with all the affordances of non-replies.
Twitter and the various Twitter-like systems enable this by treating replies as posts that can be engaged with just like the initial tweet, such as via ``retweeting'' the reply, i.e., sharing just the reply into others' feeds.
In contrast to headless Twitter, Facebook is headed.   A reply to a tweet can show up (initially) by itself in the feed of someone who follows the replier.   But if someone comments on a Facebook post, the \emph{entire post} (a head) with all its comments, not just the single comment, shows up in their friends' feeds. 
 The system has a built in assumption that heads are necessary to provide context for the comments under them.

Some systems are hybrid.  TikTok's ``Duet'' or ``Stitch'' feature, which permits quoting or embedding another video in one's own, constitutes a headless design as the ``reply'' becomes its own top-level video post.   However, the textual comments on TikTok videos follow the ``headed'' design.   So it is not clear which subspecies to assign to TikTok.   It is threaded under either consideration, however.

Yet another \emph{annotation} subspecies involves systems where comments or threads of conversation are attached to \emph{a specific location} in the ``head'' piece of content (\ref{dim:anchoring}).  This includes examples such as adding a comment to a selection of text in a Google Doc; annotating a song lyric on Genius or a selection of text from a web page or pdf document using Hypothes.is\footnote{\url{https://hypothes.is}} or Nota Bene~\cite{Zyto2012SuccessfulCD}; or attaching a comment to a particular timestamp in a video on Facebook Live.
The document or video is the central piece of content, and comments are threads that hang off it but are anchored to a specific passage or point in the content.

\paragraph{Flat Systems}
In contrast, flat designs do not connect different pieces of content through structured links. Instead, the design presents a stream of unconnected posts. 

In many of these systems, content organized in a flat way is presented chronologically in posting order, so that conversations can occur where users respond to content that appeared recently in time. Importantly, this referencing of other content happens in an unstructured way via dialogue and is not part of the data model of the system. There is also the rarer case of systems where content is not presented chronologically while appearing in a stream; an example includes Instagram Stories, where disparate images or clips of video from different users appear one after the other in a non-chronological order (content from the same user still appears chronologically). 

In terms of subspecies, content within streams can include \emph{discrete} content that is broken up into distinct units within the data model, such as a chat message, an image, or a recorded video, where users can view a stream of units one after the other. Another \emph{continuous} subspecies is content that is an unbroken stream of data shared between participants, such as a live audio or video stream, where dialogue and turn-taking happen in an unstructured way. It is not uncommon for there to be multiple interleaved flat streams of different media types; for instance Zoom offers a live video stream interleaved with a text chat stream in the same space. 

In some primarily flat systems, there may still be affordances for placing links between pieces of content; however, they often go unused in favor of simply relying on chronology. For instance, while systems like Slack and Discord\footnote{More recently, Discord has added a new product surface called ``forum channels'' where comments are required to be in reply threads: \url{https://discord.com/blog/forum-channels-space-for-organized-conversation}. This is similar to the threaded chat system Zulip: \url{https://zulip.com}. We would argue both these systems are examples of threaded systems by our definition.
} permit ``reply in thread,'' one often does not need to do so, as each channel is organized chronologically. 
Similarly, one can embed another user’s post within one’s post in an Instagram Story; the same is true for many chat apps such as WhatsApp and iMessage, where one may quote another's post to signal a reply. In all these cases, quoting is not emphasized or required for participation. 
Indeed, while one can attempt to carry on a threaded conversation using quoting in some of these platforms, it would be challenging.
There is also no option for threaded discourse under a Story in Instagram; instead, replies to a user’s story appear as a private direct message to that user with a reference to the story inline in the chat.

\subsubsection{Dimension 2: Where Content Comes From}

What determines which content you see when you visit the platform? Our second dimension is concerned with how content is delivered from one account to another. This includes users as both senders of content, where users have some ability in the system to designate how or to whom content is sent, as well as receivers of content, where users have some method of accessing content or determining how to locate a piece of content to which they have access.

We define three options for the From dimension:
\begin{itemize}
    \item \textit{Spaces}, where a user receives content that has been submitted to certain virtual locations, channels, or groups that the user ``joins'', as in Slack, Microsoft Teams, or group chat;
    \item \textit{Network}, where one receives content from specified friends or follow networks, as in Facebook, Twitter, and Instagram;
    \item \textit{Commons}, where the user notionally draws their content from the entire platform, as in TikTok or Pinterest.
\end{itemize}

We go into more detail below to characterize each option, mention common subspecies, and also discuss cases where platforms blend multiple options together.

\paragraph{Spaces}
In this option, content is exchanged in a distinct space with a unique location defined by a name or address. Space-based designs evoke the concept of physical rooms and buildings. Other offline analogies of physical spaces that researchers have used to describe social media include bulletin boards and town halls. In order to post to a space or access content from a space, a user must designate, navigate to, and potentially join the space's unique name, address, or location. Because they are defined by this unique location, in some platforms, spaces can persist even after everyone has left. Example platforms that utilize spaces are online news aggregators (e.g., Reddit subreddits), group chat (e.g., Slack, Discord), and conversations around individual pieces of non-social content (e.g., New York Times's article comments section, Google Docs comments).

Spaces may be open and have a location that is discoverable to the public, allowing anyone to freely find and enter the space. Spaces may also be hidden or access controlled. Spaces also often have metadata in addition to a unique name, such as a description, a set of rules or code of conduct, and individuals tasked with moderating the space that, together, give the space a unique set of norms and culture. 

For those who have access to the space, once they are at the location, they can peruse and potentially engage in conversations that occur within the space. Conversations in the space may be equally accessible to all people who have entered the space, or there may be further distinctions, e.g., \emph{subspaces} within a space, such as channels within Discord or Slack or groups within a WhatsApp community, where each subspace also has its own unique address or location within the larger space. Other systems take the physical space analogy farther; for instance, in social VR systems, after one joins a room, distance to other avatars in a 2D or 3D environment determines whether they can hear a conversation.

One instructive example is group chat: while it may be tempting to consider group chat as a network-based design since the recipients are specific accounts, group chat is properly a space-based design because all members of the group chat ``room'' receive all messages---and many group chat platforms such as WhatsApp even allow or encourage naming the group chat like a channel and assign an admin to manage membership. Similarly, we would characterize directed mailing, messaging, and calling systems (e.g., email, SMS, Zoom) to be spaces.  For example, although email delivery is typically individual to individual, suggesting a network model, when a user composes a message to a group of recipient, followup conversation typically makes use of ``Reply All'' to engage all the original recipients.  So the original sender essentially creates a space in which all follow-up messages are typically delivered.

\paragraph{Networks}
In network-based designs, content flows between people (or other entities with \emph{accounts}) who have connected to one another. Friend and follow networks are prototypical examples. In network-based designs, each account must have a unique way of identifying itself so that other users may find and connect to them, such as a unique username, address, or phone number. Typically, in such designs, users make friend or follow connections as they explore, or utilize discovery mechanisms to find accounts, either in-platform (e.g., YouTube recommendations) or off-platform (e.g., links to one's Patreon on another site). 

Traditional social network sites are the classic examples of social network delivery, as in the Facebook News Feed. 
Platforms that are centered around users subscribing to a creator's or influencer's content are also networks, such as Patreon, Substack, or YouTube.
Systems that implement ``stories'' such as Snapchat\footnote{Though Snapchat's Group chat surface would be considered a space.} and Instagram Stories are also networks as they involve broadcasting short disappearing videos and images to one's friends.

In contrast to spaces, where there are unique shared locations where people can gather, and any user in the location can access that content, networks distribute content from account to account. Importantly, this means that while spaces provide common knowledge and shared context---everyone in the space can access the content posted in the space---in networks everyone sees a different subset of content, and it is not always clear what subset others see~\cite{gilbert2012designing}.

\paragraph{Commons}
Some designs feed content to users without requiring any following, joining, or subscription at all.  Instead, users receive content from everyone and anyone on the platform. Algorithmic ranking is often central to commons-based designs. The most well known commons design as of writing is the ``For You'' feed as popularized by TikTok, but this also includes Instagram's ``Explore'', Snapchat's ``Discover'', Reddit's ``Best'' or logged-out surface, and the Pinterest feed. To the extent that users' profiles are the main content to be shared in a network like Tinder, or that video streams are matched together randomly on ChatRoulette, we may also view them as commons-based designs. The algorithms used for distribution can be complex, with many factors and methods of inference, or simple, e.g., random or chronological algorithms.

\paragraph{\textit{Blended Delivery Mechanisms.}} It is not uncommon for platforms to blend multiple delivery mechanisms (our ``From'' dimension) for the same content or in the same product surface, particularly as algorithmic feeds have become more popular in recent years. In each case, we identify a primary method of delivery by examining the defaults in the system. 

There are many systems that combine both social networks and spaces. Examples of platforms where a follow graph and a social network feed has been tacked onto a space-oriented system include systems such as Quora, Github, and Reddit. In these cases, the primary orientation is around spaces but users can additionally follow individuals and see content from those individuals in a home feed. 
Another example that is less clear-cut is Mastodon, where one’s “Home” feed consists of content from the people one follows (a social network), while one’s “Local” feed consists of content from all the people in one’s local Mastodon instance, whether or not one explicitly follows them (a space). Despite having distinct elements of both, we would characterize Mastodon as having more emphasis on a social network model of delivering content due to the defaults provided by its current main clients (i.e., less emphasis on local content, more emphasis on one’s followed content). 

Many platforms also blend commons designs with other delivery mechanisms. 
As of this writing, Twitter's For You page is 50\% centered around follows and 50\% content from the commons based on a ranking algorithm. TikTok's For You page is more skewed---principally a commons design but somewhat influenced by the follow network. Many social networks are beginning to augment their social network approach with the commons, claiming that this delivery mechanism increases engagement and content quality. For instance, Meta's Threads app recently launched in mid-2023, with more of a commons-based delivery as it seeks to engage users who are still building a following graph.

Some platforms combine elements of all three delivery mechanisms. Reddit has both spaces (subreddits) to which users may visit and subscribe, but also a home feed showing content from subreddits to which users have not subscribed and content from users they follow; however, in this case, we would still say Reddit is still primarily oriented around spaces.

\subsubsection{Ideal Types for Each Cell in the 2x3 Model}
To help conceptualize the Form-From model, we find that it can be helpful to consider Weberian ``ideal types''~\cite{weber2017methodology} for each cell. In Sociology, ideal types are an approach for building abstractions and theory in the face of the complexities of the real world by creating hypothetical but clear examples. In our case:
\begin{itemize}
    \item Flat spaces: a chat room with chats appearing in a traditional chronological order
    \item Threaded spaces: an online forum, with a top-level post anchoring each thread and replies inside the thread
    \item Flat network: a social network ``wall'' where friends can leave posts that render in a reverse-chronological order (e.g., the original Facebook Wall)
    \item Threaded network: a social network feed that features posts from followees, where each post can receive comments and reshares
    \item Flat commons: a site for viewing algorithmically curated content, without the ability to reply or follow accounts
    \item Threaded commons: a site for viewing and replying to algorithmically curated content, without the ability to follow accounts
\end{itemize}
Ideal types such as these can help us characterize and categorize actually-existing systems, which are rarely so clear-cut.

\begin{table}
\small
\begin{tabular}{|l|p{10cm}|}
 \hline
\textbf{Category} & \textbf{Example Axes} \\ \hline 
Discussion structuring & 
How are contributions grouped (individual posts \textit{a la} Twitter, stages or phases of discussion \textit{a la} deliberation platforms, threads \textit{a la} forums, or time bounded blocks \textit{a la} Zoom meetings)? Is there a single channel, or multiple channels? How are contributions anchored (to posts, to timestamps, or to a spatial coordinate system? \\ \hline
Delivery &
How are contributions ordered (chronologically, algorithmically personalized, voted upon by the group, editorial fiat)? What can the user block (people, topics, replies to a post, by score, or nothing)? \\ \hline
Temporality &
Does the platform require synchronous interaction? How durable are the contributions (enforced ephemeral \textit{a la} 4chan, default ephemeral \textit{a la} Zoom, default archived \textit{a la} Slack, or enforced archived \textit{a la} Wikipedia)? \\ \hline
Audience & 
What is the maximal audience one can reach (a bounded group, a networked neighborhood, or public)? What is the typical audience one reaches (an unknown subgroup of platform members \textit{a la} Twitter, entire membership of the group \textit{a la} email lists, public within the platform, or a pair or small group \textit{a la} WhatsApp)? \\ \hline
Membership &
What are the criteria for joining (geofencing, referral from a current member, payment, identity characteristics, organizational membership, open)? What user roles exist (member, moderator, administrator)? \\ \hline 
Moderation &
Who is empowered to make decisions (members, paid moderators, volunteer moderators, platform developers)? Are moderation actions transparent (invisible to members, visible to members)?
\\ \hline 
Governance &
How do members gain access to power (elections, reputation/karma systems, appointment by existing elites)? How are changes to governance determined (direct democracy, elite members decide, platform operators decide)? \\ \hline 
Extensibility &
Are bots allowed (conversational bots, operational support bots)? Is there an available API (to extract data, to create bots, to contribute to discussion, to federate with other instances)? \\ \hline
User and Group Signaling &
How does a user convey themselves (profile image, display name, real name, avatar)? Are there availability indicators for users (here/away status, typing (``...'') indicators, location markers)? \\ \hline 
Content &
What types of content are allowable (text, image, audio, video, avatar, drawing, emoji, etc.)? What kinds of authoring tools are available (AI-MC, filters/lenses, overlays, accessibility hooks, remixing/stitching)? \\ \hline

\end{tabular}
\caption{The ten expanded categories of our Form-From model, and example axes within each category.}
\label{tab:10dim}
\end{table}

\subsection{10-category expanded model}
The Form-From model captures two principal dimensions of variation across social media designs. However, a simple model cannot always capture dimensions of difference that matter to a particular system. For example, Mastodon, Bluesky, and Twitter all fall into the Threads + Network cell of the From-Form model, but each of them can also be described by important differences. 
 For instance, Mastodon and Twitter differ strongly in their models of governance and extensibility, and all three differ strongly in their moderation strategy (Twitter's is centralized and increasingly hands off, Mastodon is decentralized to server administrators, and Bluesky places it in the hands of end users through middleware).

This section describes the 10-category expanded model. This model hierarchically encompasses all 62 dimensions in the full model. 
Depending on the practitioner or researcher's analytical goals, they might select different axes than the two we identified above. 
Form-From captures axes of substantial explanatory variation for questions of interest, but
for some applications, other axes such as Temporality or Audience may be the more relevant axes of differentiation.

\paragraph{Discussion Structuring}
This category most directly aligns with the Form dimension described above but goes into more detail to characterize the different structures or data objects that social media systems might take. 
In particular, whether the system groups utterances into distinct ``post'' objects or not and whether these posts can be linked to each other as chained reply ``threads'' or not are fundamental differences in discussion structures.
Other dimensions include: What depth of threaded replies does the platform enable? Reddit has infinitely deep reply trees, while Facebook displays only two levels of replies. Are some posts distinguished from typical or generic posts? Q\&A sites like Quora have questions and answers, while Pro-Con systems like Kialo have initial statements, followed by pros and cons.

\paragraph{Delivery}
The delivery category most directly aligns with the From dimension described above but is more broadly interested in different aspects of how content gets delivered to users and the degree to which users can specify aspects of what gets delivered to them (or more specifically, what they can see, find, or be made aware of). Examples of dimensions we cover include: Is there a feed?  If so, how are posts ranked in it, and how are replies to a post ranked? Messaging platforms like WhatsApp often partition content by spaces and have chronological ranking within each space, while most social networks have a personalized algorithm based on engagement metrics. Can users decide what kinds of posts they want to see (i.e., subscription) or what kinds of posts they \textit{don't} want to see (i.e., muting)?

\paragraph{Temporality}
The temporality category parallels the time axis in Johansen's time-space model but with additional nuances. Social media systems can differ based on the assumptions they make around synchronous interaction, as well as their retention policies. The temporality category contains dimensions such as: Does the platform require synchronous interaction? Platforms meant for real-time conversations, such as Clubhouse and Zoom, do while many others such as Soundcloud and YouTube do not. Likewise, how ephemeral is the content? Zoom chats and 4chan posts disappear, while traditional internet forum submissions are retained indefinitely.

\paragraph{Audience}
Platforms often give users options about the audience they wish to reach with their posts, but their defaults are also meaningfully different. For example, for the past decade Facebook's default audience has been friends-only, whereas Mastodon, Twitter, and Bluesky are world-public by default. This category contains dimensions such as: What is the maximal audience one can reach (a bounded group, a networked neighborhood, or public)? What is the typical audience one reaches? For example, an email list or WhatsApp group typically reaches \emph{all} and \emph{only} members of the group, whereas platforms such as Twitter and TikTok typically reach an unknown subgroup of platform members.

\paragraph{Membership}
Platforms sometimes define themselves by who is present, or how people earn membership. For example, the Metafilter community famously required a nominal \$5 fee to join, in large part to keep the community self-selecting to those who cared enough about it to put in a little money. Others, such as YikYak, require geofenced physical presence to join. The membership category contains dimensions such as: What are the criteria for joining? (Options include geofencing, referral from a current member, payment, identity characteristics, organizational membership, and no requirement.) Or, what user roles exist (e.g., member, moderator, administrator)?

\paragraph{Moderation}
Moderation can be a differentiator between platforms: as Tarleton Gillespie argues, “moderation is, in many ways, the commodity that platforms offer”~\cite{gillespie2018custodians}. The moderation category addresses questions about how content moderation is performed on the platform. For example, this category contains a dimension naming non-exclusive options for who is empowered to make moderation decisions (e.g., members, paid moderators, volunteer moderators, platform developers).

\paragraph{Governance}
Who makes decisions about how a platform operates?
Governance can lead to differences in designs and socio-technical outcomes, for example whether a platform provides tools and autonomy to groups or subsets of its users. How do members gain access to power:  elections, reputation/karma, or appointment by existing elites? How are changes to governance determined: direct democracy, elite members decide, or platform operators decide?

\paragraph{Extensibility}
Platforms can be extremely customizable, if the developer allows it. Discord, for example, is known for its variety of community-developed extensions, whereas Microsoft Teams is more a walled garden. Extensibility includes axes such as whether bots are allowed, including both conversational bots and operational support bots. Reddit, for example, commonly allows bots, whereas Twitter has been increasingly cracking down on bot accounts. Likewise, is there an available API to extract data or to contribute to the platform from a custom client?

\paragraph{Signaling: Users and Groups}
How do accounts and groups signal their identity and availability to others? Reddit allows groups to deeply customize the look and feel of their spaces, whereas Facebook Groups are much more restrained. The Signaling category articulates design decisions platforms make that shape how we project our identity and behavior out to others on the platform, focused on aspects aside from the main content itself. For example, how does a user present themselves (profile image, display name, real name, avatar)? Mastodon profiles focus on an image, display name and avatar, whereas Facebook has a real-name policy, and 4chan offers no facilities for a profile or continuous identity at all. Likewise, platforms offer different availability indicators: Slack uses colored circles to convey a user's presence or away status, but iMessage does not; iMessage conveys typing (``...'') indicators, whereas WhatsApp does not. 

\paragraph{Content}
To the mass media, the type of content often defines a platform. Instagram was ``the photo platform'', while Twitter was the ``microblogging platform'' and TikTok or Vine were the ``short-form video platforms''.  While we would argue that the content type is less determinative than these portrayals might imply, it still matters. What subset of content types are allowable on the platform: text, images, audio, videos, avatars, drawings, emojis? Platforms may also provide facilities for certain types of content, for example Instagram's filters and TikTok's stitch creation tool.

\subsection{Full model: 62 dimensions}
Our full model consists of 62 dimensions, the representation of the complete design space. Appendix~\ref{adx:model} contains the full set of 62 dimensions that comprise the ten categories above. Sixty-two is not a magic number, and we expect that these dimensions, and their options/values, will need to evolve as social media design does.

The full list of dimensions should be helpful as a checklist for designers: every social media design must make decisions on each dimension. Not making a decision means that the design does make a de-facto decision to go with the typical or default value of the dimension. (For example, by default, platforms permanently keep archives.) However, defaulting can lead to mismatched expectations between users and platforms, and give rise to conflict later. So, we advocate that one practical use of our full 62-dimension model is as a design checklist for designers in the early stages of design.
The full model is also useful for deep-dive inspections of closely related systems. For example, for research focused on social identity~\cite{seering2018applications}, comparative analyses across platform design decisions related to Membership and User and Group Signaling may be useful: What are the criteria for joining the platform or a group? Can a single person hold multiple accounts?

In the following, we walk through a deep dive into just one of the dimensions to explain each of the options, how we arrived at those options, and decisions we made along the way. 
This serves as an example of the  reasoning we used to define the dimensions in our 62-dimension model.

\subsubsection{Dimension: Main Location of Discussion} 
\label{sec:62dim-deepdive}
This dimension lives under our Delivery category and focuses on high level questions about where social content resides, how it arrives there, and how a user accesses that content on a platform. Given the name of the dimension, spatial metaphors highly inform the options.  Note this dimension can take on multiple values.

For many systems, the way to access content is simply to navigate to a single address, or a \textit{center}, where all discourse is hosted and available. Further dimensions in the Delivery category address how to find specific content, or be notified of or subscribe to content, or in what order content is presented. All of these questions have significant importance when a system is a \textit{center}, and thus has a large body of undifferentiated content in one place that needs further ways to characterize and distinguish content. Indeed, most systems with this option operate some sort of feed for readers.

A second option we observed for the dimension Main Location is \textit{rooms}---a special case of spaces in the From dimension---where instead of one center where all discussion takes place, the space is first divided further into distinct addressable partitions, where users can enter the room to peruse content or post to it. Examples include Slack channels and Facebook Groups. One debate we had around this option is whether hashtags as popularized on Twitter, or tags in general, should count as ``rooms''. Eventually we decided that while tags and rooms have some similar practical implications, they also have important differences, namely that while a piece of content may have many tags, they typically only have one main location (though can possibly be reposted or shared elsewhere). This discussion is also what led us to title this dimension ``\textit{Main} Location of Discussion'', recognizing that the same content can reside in different places. Instead, tags and hashtags appear in separate dimensions in this category such as  \textit{Subscription} (\ref{dim:subscription}) and \textit{Findability} (\ref{dim:findability}).
A second debate we had was whether profile pages on systems such as Facebook and TikTok could be thought of as rooms, or whether these systems are centers. We eventually agreed that while one's profile page has some properties in common with rooms, namely a unique address where content can be found, we still classify these systems as center-based because their design is not primarily organized around navigating to profile pages in order to see content. However, a caveat is that early versions of some social networking systems like the Facebook Wall before the feed and their predecessors such as Blogger were room-like---one had to navigate to someone's blog or wall to access any content.

Next, we separated out \textit{cliques}, which are like rooms, but instead of having an addressable name, cliques are distinguished by the membership of the shared space.
An example of this distinction is found in Slack, where there can be channels with names like ``\#general'', but there are also multi-party direct messages, which have no name but are distinguished by the participants. An important difference between how rooms and cliques work is shared history---as users get added or removed from a channel, the historical content stays intact, much like one would expect of objects in a physical room where people can walk in and out. Indeed, a room still exists even if it is entirely empty of people. In a clique, however, if one adds a new user to a multi-party message, it is now a new clique with its own distinct history. We have found this subtle distinction leads to some interesting and unexpected behavior in different systems. For instance, we found through investigation that Facebook Messenger's multi-party messages are closer to rooms than cliques, e.g., members of a multi-party message can add new people to the chat who can see history from before they were there, and members can also rename the shared chat as well as appoint admins.  Facebook's user experience confuses this issue: if a user composes a message to the group by listing their names, and they have messaged the exact same group previously, then Facebook drops them into the preexisting group channel where previous messages can be seen.   This creates the illusion that communication is based on the clique.  However, if the group's old room gets renamed, then messaging the same group results in the creation of a new, different channel for the new conversation.
From our investigation, almost all group chat systems actually are designed more like rooms.

The final option is \textit{mailboxes}, where there is no shared location for discussion but all content gets delivered to personal inboxes. Email is the canonical example, where even if an email is removed from one participant's inbox, it is still available on all other participants' inboxes, as they each have their own copy. In addition, each user can choose to customize and use their desired email client as opposed to the shared experience of a common center or room.
We discussed among ourselves how technically, most email systems are now hosted on a centralized service via the cloud as opposed to physically on a client device. Nevertheless the design of some email systems is still reminiscent of physical mail deliveries. There are also increasingly systems that would traditionally act like mailboxes, such as direct messaging, that now have room features in their designs.

Our iterative discussions helped us isolate the important dimensions that made their way to the 2x3 model. In the case of this dimension, we determined that the design distinctions today between rooms, cliques, and mailboxes are overall subtle, so when we went to summarize our dimensions in our higher level 2x3 model, all of the systems with these values actually end up in the \textit{Spaces} option of the ``From'' dimension. In contrast, systems that have one single center end up not having much of a spatial metaphor, due to the primacy of algorithms curating a central feed reducing the need for much navigation. Instead, these systems
became spread out between the options of \textit{Network} and \textit{Commons} according to their placement on lower level dimensions such as \textit{Ranking of Posts} (\ref{dim:ranking-of-posts}), or in what order posts appear, combined with \textit{Maximal Audience} (\ref{dim:maximal-audience}), or to what posts do users have access. \newline

As can be seen, while some dimensions were relatively straightforward, others such as Main Location required significant discussion and investigation. Our discussions also heavily influenced how we formulated our 10 categories and the 2x3 model. While a full accounting of each of the 62 dimensions would be far too long to delve into here, this deep dive provides an example of  how we came to the possible options for each dimension, how we investigated their values in different platforms, and the debates we had in shaping each dimension and relating it to other dimensions.

\section{Application and Extensions of the Design Space}

What can we do now that we have a design space? What questions can we answer? We apply our model to a series of case studies in both product and research systems.

\subsection{Evolution of Social Media Over Time}

The Form-From model allows us to explore how social media designs have evolved over time. What path took the field from the early days of BBS to today's platforms?

To trace this path over time, we began with a timeline of major social media platforms sourced from Wikipedia.\footnote{\url{https://en.wikipedia.org/wiki/Timeline_of_social_media}} This timeline named 77 social media platforms and the year of each platform's launch. We then deductively applied the Form-From model to each of these platforms. Since the platform designs evolved over time, we applied a From-Form code to the design as of the launch of the system: for example, when Facebook launched, it was a flat-network system rather than a thread-network system, since users could post on each others' walls but there was no functionality to reply or comment on each others' posts.   
Our method involved a single author coding all 77 systems in a first round through historical and archival research on the web, then the remaining coauthors reviewing the codes in a second pass with a discussion to resolve any disagreements. A full table reporting the systems and codes is available in Appendix~\ref{apx:history}. This historical analysis was reliant on both primary and secondary sources available on the internet about these platforms, and so we stress that future work may refine our work here.

\begin{figure}
    \centering
    \includegraphics[width=0.8\textwidth]{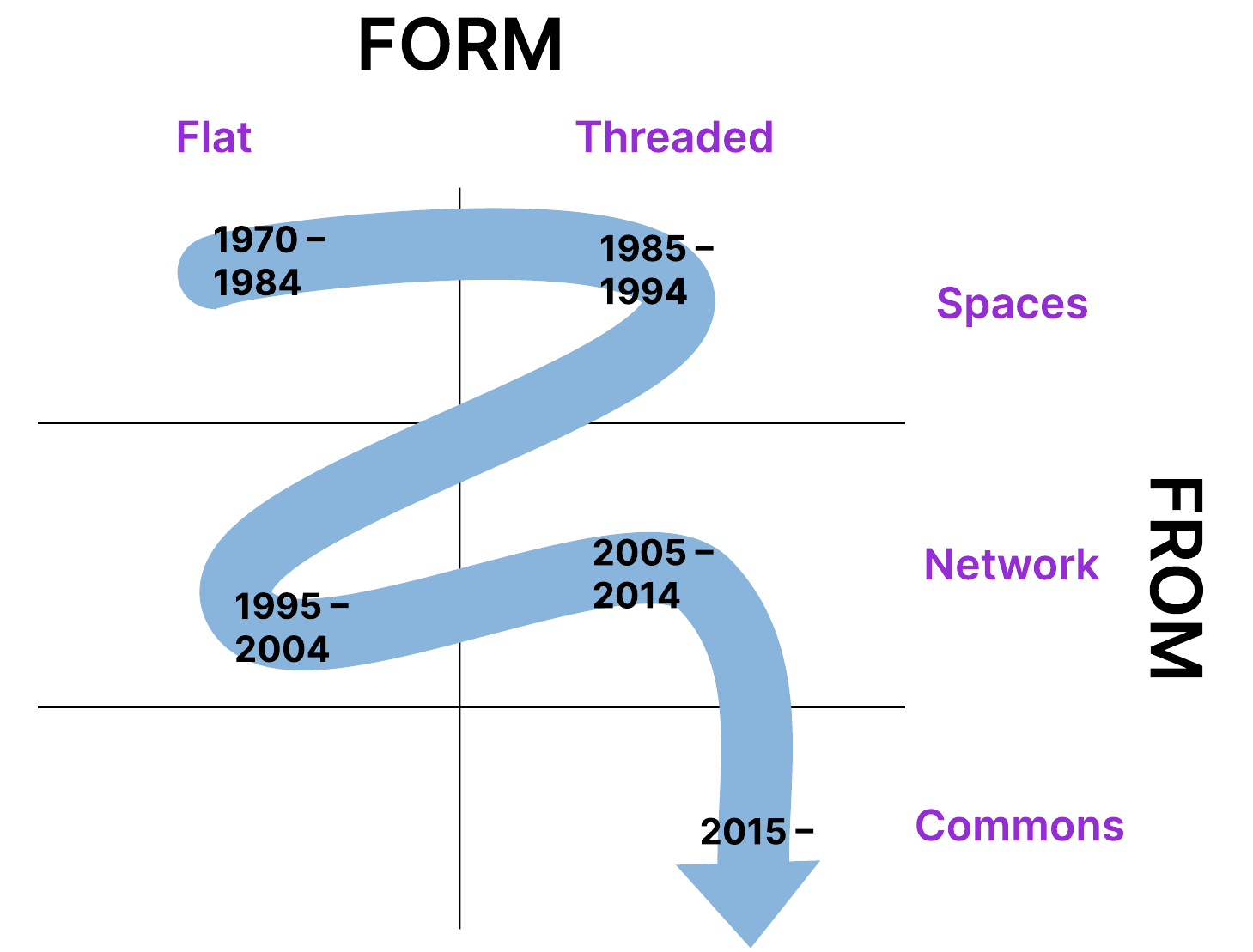}
    \caption{Social media systems followed a pattern over time of moving from flat spaces (e.g., Talkomatic), to threaded spaces (e.g., Usenet), to flat networks (e.g., the original Facebook), to threaded networks (e.g., Reddit), to threaded commons (e.g., TikTok). This path does not describe every system---it describes an overall pattern.}
    \label{fig:overtime}
\end{figure}

We can characterize social media's evolution over time through five eras. Figure~\ref{fig:overtime} lays out the path of this evolution, and we discuss example systems below.

\paragraph{1970--1984: ``Flatspace''}
The 1970s and early 1980s were firmly rooted in space-based designs, and especially flat designs. PLATO Notes (1974) was the only listed system in this period to feature threading. Even email and Usenet, which were not included in the Wikipedia dataset but qualify according to our definition, did not begin to make widespread use of threading until the late 1980s: until then, typically any context had to be added manually through quoting text, and messages were interleaved in a chronological stream rather than grouped as threads. Early flat, space-based systems included Talkomatic (1973) and BBS (1980).

\paragraph{1985--1994: Threads Help Manage Scale}
In the earliest days, threading was less necessary because communication volume was relatively small. However, beginning in the late 1980s and early 1990s, as these platforms grew, threading became more popular to manage the load. Eudora, the email client that would introduce email threading to a large audience (in part through the use of a ``In-Reply-To'' mail header), launched in 1988, and the \texttt{trn} threaded Usenet reader was launched in 1993. This era gave rise to the model of the threaded discussion forum that we are familiar with.

\paragraph{1995--2004: The Rise of Messaging and Networks}
Beginning in the 1990s, social networks as we conceive them began to arise. However, the first step was a snap back to flat space designs through the rise of instant messaging systems, including AOL Instant Messenger~(1996), ICQ~(1996), Yahoo! Messenger~(1999), and MSN messenger~(1999). These platforms could be flat because small group conversations are much smaller in volume and easily tracked without threading (especially when synchronous).

Starting in 2000, network-based platforms began a rapid ascent. In fact, while network-based systems were relatively rare up through 2001, ten of the eleven systems introduced in 2002-2004 were network based. Early network systems included Classmates.com~(1995), SixDegrees.com~(1996), and LiveJournal~(1999). Then, there was a large crush of familiar network-based platforms launching starting after 2000: Friendster~(2002), LinkedIn~(2003), MySpace~(2003), and Facebook~(2004). Each of these were still flat designs, however, as (for example) friends' walls on Facebook were a flat stream of posts and could not support commenting or linking.

\paragraph{2005--2014: Networks Rediscover Threads}
Through this entire period of 1970--2004, flat was the dominant system design. While there were occasional thread-based platforms such as PLATO Notes~(1974), Bolt.com~(1996), and LiveJournal~(1996), only 22\% of the systems coded in 1970-2004 were thread-based. But then, starting in 2005, and roughly contemporaneously with the launch of the first Facebook News Feed, social media platforms began introducing threading. Quickly, nearly all new platforms switched from flat to threads: 80\% of systems 2005-2023 are threaded. This inversion from 20\% thread / 80\% flat to 80\% thread / 20\% flat happened quite suddenly in 2005, starting with Reddit~(2005), YouTube~(2005), Twitter~(2006), and Tumblr~(2007), and continuing with platforms such as Pinterest~(2010), Instagram~(2010), Snapchat~(2011), and Musical.ly (the predecessor of TikTok---2014). Flat-based outliers continued to exist, but they were the minority.

\paragraph{2015--current: Algorithmic Commons}
While the systems up to this point required users to explicitly specify what they wanted to receive, progress in machine learning eventually reached the point where it became feasible for systems to learn user preferences implicitly and select content for them from the commons. There are only four platforms in our dataset that delivered from the commons at launch. Two of them occurred before 2015, which we are marking the start of the commons era: Pinterest in 2010, and Tinder in 2012. However, it is in 2017 that TikTok brought to the fore the idea of a platform sourcing content from the commons, and algorithms became a major focus in media and research---so we place the start of the era around here, at 2015. However, six years after the launch of TikTok, only Meta's new platform Threads in the dataset we would argue is primarily algorithmic---perhaps users still want to maintain some degree of explicit control over what they see. Other platforms since 2015 remain purely or predominantly space and network based, including Discord (2015), Mastodon (2016), BeReal (2020), Clubhouse (2020), Gab (2016), Parler (2018), and Truth Social (2021).

Instead, many platforms remained in their existing cell, but fused elements of the commons into their existing design. For example, Twitter's For You feed is now 50\% content from followed accounts and 50\% content recommended from the commons.\footnote{\url{https://blog.twitter.com/engineering/en_us/topics/open-source/2023/twitter-recommendation-algorithm}} Likewise, Instagram increased the proportion of posts from the broader platform in its feed, then reduced it again after backlash.\footnote{\url{https://www.theverge.com/2022/7/28/23282682/instagram-rollback-tiktok-feed-recommendations-interview-adam-mosseri}} So, despite multiple proclamations of the death of network-based designs,\footnote{\url{https://www.newyorker.com/culture/cultural-comment/tiktok-and-the-fall-of-the-social-media-giants}, \url{https://www.nytimes.com/2023/04/19/technology/personaltech/tiktok-twitter-facebook-social.html}, \url{https://www.axios.com/2022/07/25/sunset-social-network-facebook-tiktok}} few platforms have fully converted to a commons-based model even years after the rise of TikTok.

\subsection{Reflecting on Social Media Designs using Form-From}

\subsubsection{Does Scaling Drive Movement in the Design Space?}  
For a small group engaged in just a single conversation, a flat space design is certainly the simplest.   The other cells of our matrix can all be seen as additional complexity introduced to cope with scale.  As the number of conversations grows too large, spaces allow a participant to pay attention only to conversations within sub-spaces of interest (e.g., channels).   Networks let participants focus attention on people they care to hear from.   Threads partition overlapping conversations so that a participant can attend to one at a time in its entirety instead of being forced to keep all of them in mind at once.   Even if we get to the point of algorithms doing a perfect job of selecting the content we should see, selecting specific spaces, following people, and attending to specific threads still provide valuable organizing infrastructure to help people think about one thing at a time.   

While these changes alleviate the pressures of scale, they also create unfamiliar social configurations that result in unexpected behavior and eventually new social norms and designs. For instance, Marwick and boyd~\cite{marwickContexCollapse} were the first to point to context collapse in social media: the situation where a post aimed at one of a user's multiple audiences was delivered to all of them.  Context collapse tracks the move of social platforms from spaces to networks and the commons.  Spaces align well to contexts, as a user aiming to reach a particular audience can post to a space, where the audience is a defined set.   The move to networks and commons designs eliminates that easy path for posters to target their posts to an appropriate audience.

As content volume continues to scale up, we can compose multiple ``layers'' of delivery designs---which reduce content volume---in order to cope.   Slack's channels within workspaces are a spaces-in-spaces design.  Another alternative to handling a space---such as a subreddit---that got too big would be to use a network design within it, delivering to a user only the content in that space that was posted by someone they follow. This combination of space and network designs would protect participants from context collapse, as they would only see content from a given friend \emph{on a specifically chosen topic}. 
Instagram's Close Friends list could be seen as a network-in-network design to further filter down from a large following list.
As another example, we see Facebook applying algorithms---which naturally apply to the commons model---to filter comments on a single post down to only the ``most relevant,'' as opposed to showing all replies.

\subsubsection{Headed Content as Spaces}
We've discussed the ``headed'' subspecies of the threaded form, in which comment threads hang off some initial privileged type of content.   This is quite reminiscent of the way content is delivered from spaces, if we simply imagine that each head defines a space whose participants are the people interested in discussing that particular piece of content.  This creates some design opportunities which have not yet been extensively explored: some YouTube videos have as many as 500,000 comments, far too many for any individual to consume.   Considering that video as a space, one could imagine applying alternative From models to the content of the space: perhaps a network model, in which users only see comments from people they follow or a partition of the comments according to multiple defined subspaces.  Or, treating the space as a commons and using algorithmic techniques to select comments to show (as mentioned above, Facebook offers this method as ``most relevant'' comments). In addition, many of the moderation features that are common in space-based designs, such as approving members, appointing moderators, and removing or promoting comments, could be introduced. For instance, we have seen some networked platforms introduce the ability to moderate replies to one's ``head'' post, such as Twitter's Hide Reply feature in 2019.\footnote{\url{https://techcrunch.com/2019/11/21/twitter-rolls-out-its-hide-replies-feature-to-all-users-worldwide/amp/}}

\subsubsection{Abandoning and Reclaiming User Control}  Earlier systems used spaces and then network designs; it is only recently that the commons
design has emerged.  One possible reason for this timing is that the commons demands highly sophisticated algorithms that have only recently been developed.  On a large social platform, there has to be some way to deliver a tractably small subset of messages or posts to an interested consumer.  Historically, limiting delivery to a small, user-selected collection of spaces or people of interest could do that.  Today's sophisticated machine learning algorithms instead offer the possibility of learning a user's preferences and selecting a small amount of content suited to that user from the pool of posts from the commons.  This approach could not succeed until the algorithms developed sufficient precision to ensure that most of what they delivered was interesting to a user.  It also works best in situations such as entertainment where no one piece of content is essential to deliver to a user, but any sufficiently interesting content will suffice.   Using algorithms instead of spaces or networks means that users cede control over what they will receive to those algorithms.   
However, as we've seen with the backlash to Instagram's change to their feed algorithm and with the recent interest in decentralized Twitter-like systems like Mastodon and Bluesky, users are also interested in control. 
In the future, we may see new designs that aim to innovate in both providing users some control, yet leverages the power of algorithms to reduce user labor in articulating their preferences~\cite{feng2024mapping}. One example is Bluesky's recent introduction of custom feeds, where users can create their own algorithmic feeds and choose from a suite of feeds published by others.\footnote{\url{https://www.theverge.com/2023/5/26/23739174/bluesky-custom-feeds-algorithms-twitter-alternative}}

\subsection{Identifying Design Patterns within Form-From Cells}
\label{sec:design-patterns}

As first articulated by \citet{alexander1977pattern} and then translated into HCI by \citet{van2007design}, a design pattern is a design that is commonly used to solve a given problem. When we find a cluster of well-known systems in the design space with similar values for many of the dimensions, we can consider it to be a design pattern.

Design patterns can arise when certain dimension values work better with or directly imply other dimension values. For example in the case of group chat systems with multiple channels (e.g., Slack, Discord, Microsoft Teams), real-time awareness in the form of signaling of real-time user activity and status is useful for enabling synchronous conversation, similar to other chat and IM systems. A number of other designs come bundled with the high level choice of supporting synchronous messaging, such as emphasizing short text messages and chronological sorting. However, unlike group chat systems for smaller groups such as GroupMe or Snapchat, systems like Slack and Discord can scale to larger groups through enabling many parallel conversations via channels. 
Hence, it is not surprising to see WhatsApp recently introducing a Communities surface enabling over 20,000 members with the feature of multiple groups, as a contrast to their Groups surface with no subgroups and a limit of 1,024 members.\footnote{\url{https://blog.whatsapp.com/communities-now-available}} 

Other common design patterns that we observe within specific cells of the Form-From model:
\begin{itemize}
    \item Threaded Spaces: upvote/downvote ranking feedback (e.g., Reddit, Quora); overview pages allowing users to skim top-level posts or recently updated threads (e.g., phpBB, 4chan);
    \item Flat Network: to post content to a person on the network, visit the person's page rather than a centralized posting interface (e.g., MySpace, early Facebook Wall, early LinkedIn);
    \item Threaded Network: common social network design patterns such as likes/favorites, shares, comments; vertical scrolling feed with intermixed content (e.g., today's Facebook NewsFeed, Twitter);
    \item Threaded Commons: content from the entire platform is algorithmically recommended based on user feedback, and users are encouraged to comment on the content (e.g., TikTok).
\end{itemize}
Many of these design patterns arise commonly within a given cell but would be nonsensical in other cells, supporting the internal consistency of our model.

\section{Discussion}

\subsection{Implications for Designers and Researchers}
While our main effort was to develop theory, we aim for our models to provide levers for practitioners and designers as well. One strategy for applying our model is to find design inspiration and explore alternatives. Designers might start with their proposed new design, then identify its attributes on our model's dimensions. Having thus determined the location ``coordinates'' of their system in our design space, the designer can then use our model to identify other systems that are located in the same or nearby cells in the model. By doing so, they can begin to observe related design patterns, understand how those patterns played out in actual usage, and pick and choose elements of those designs that they want to consider for their own design.

Our larger models can also serve in a practical capacity as checklists. Every new system must implement a value for essentially every dimension in our model. If we don't design with intent, we default. In other words, it is far better that designers walk through the model's dimensions and think critically, up front, about their design decisions, than to make them implicitly or inherit them unthinkingly from whatever technical framework they are using.

As researchers, we investigate social media design, yet there is still much to learn. Our taxonomy may be helpful for mapping out productive lines of research. For example, what influence do individual dimensions, or intersections of these dimensions, have on the sociotechnical outcomes of platforms and communities? We view these models as launchpads for research agendas to better underpin our understanding of social computing design.

We also hope that this effort helps sharpen our language and thinking around social media systems. There has been an inherent fuzziness to terms such as ``algorithmic feed'', because these feeds can combine elements of several different dimensions. However, by staking out whether we are talking about an algorithm being run over a commons, or over a network, or (as in Twitter) a 50--50 combination of both, we can have more focused discussions. Going forward, these dimensions can also give rise, we hope, to more robust design pattern languages as well.

\subsection{Current Limitations and Future Work}

Like all researchers, we would like to believe that our work is ``future proof''---but we doubt it is.  The goal for a scientific model is an understanding that is true regardless of historical circumstance.  An engineering model, particularly a Simonian model, must change as new devices or systems are constructed, for example because of new capabilities in the environment (e.g., recently, the low cost of distributed cloud storage or ubiquitous networking).  Thus future work is likely.  We foresee five lines of future work we believe will be profitable.

 \textit{Adding detail to dimensions.}  We believe over time, new systems and system capabilities will require additional nuance and detail for some dimensions.  One example of this is with algorithms.  One of our challenge was how to conceptualize algorithmic ranking as part of our design space, as algorithms have become more and more salient parts of the social media experience. An earlier version of the model replaced ``commons'' with ``algorithm''. However, essentially every modern threaded system, from threaded spaces such as Reddit, to threaded networks such as Twitter, to threaded commons such as TikTok, uses algorithmic ranking. Ultimately, we decided to focus the From axis on the potential sources of content, and view algorithmic ranking as orthogonal. While not including ``algorithm'' in the top-level model downplays the visibility of algorithmic ranking, we argue that the From axis better explains existing variation in platform design.

However, we expect to see the trend of algorithmic commons continue until the use of algorithms to make content visible or detectable to users will be a component of many more platforms. At that point, it may become important to distinguish between different kinds of algorithmic delivery, much as we’ve needed new terms beyond asynchronous and synchronous to understand how platforms use time today. In our full 62-dimension model, under the Delivery and Audience categories, we start to separate out how algorithms are used---whether for determining audiences versus ranking of content versus ranking of replies. But in the future, additional consideration may be needed to distinguish different kinds of algorithms, or the different kinds of signals they incorporate.

\textit{New dimensions.}  We do not claim that our model captures all variations in platform design. A Simonian design space aspires to do so, and so our full 62-dimension model is a useful effort.  Nonetheless, the long tail of designs argues that our 62-dimension model will be incomplete for some analytical goals, and that some of the categorical options for its dimensions can be refined. We welcome this improvement.

\textit{Understanding the relationships between dimensions.}  Our design space was defined as a hierarchy (i.e., a tree).  Our models all have a tree structure, which is useful for taxonomic purposes and allowed the problem to be tractable.  Trees, however, cannot easily represent necessary or preferred relationships between dimensions.  Now that we have delineated the models, future work may consider relationships between attributes (dimensions) and even values, thus forming a network structure.  This will be important for considering subspaces of the design space for design patterns. It will also be important for fully examining dynamics in social media systems, such how the social networks form \cite{jacobs2015assembling}.

\textit{Design patterns and operators.}  Card, Mackinlay, and Robinson argued for the inclusion of operators in a design space analysis~\cite{card1990design}. For them, operators provided for the composition of applications of the design space.  Our work has examined the analytical dimensions upon which applications are constructed; future work should consider how they are used together in compositions. This would allow one to broaden from examining one product surface at a time to consider how platforms combine multiple points in the design space together into one experience.  The combination of multiple dimensions with specific values, along with the composition operators, would lead to design patterns for social media systems, as we touched upon in Section~\ref{sec:design-patterns}.

As mentioned in the previous subsection, future work could also consider complementing our inductive effort with quantitative studies. For example, are there correlative or even causal relationships between design decisions in one part of the design space and design decisions in others? For example, is blocking plausible for flat-space systems? Comparative analysis across platforms or via natural experiments might likewise answer, what are the impacts of each decision on user behavior or emergent platform norms? For example, do differences in governance lead to differences in the impacts of disputes on community social capital?  Empirically-based hypothesis-testing studies could lead to a stronger understanding of how these dimensions could best be combined.

\textit{Sociotechnical dimensions.}  In order to make the modeling problem tractable, we simplified it by examining only technical characteristics of what are necessarily socio-technical systems. 
A  future line of work could try to also include dimensions of social behavior, i.e., not just how systems are designed but how they are appropriated by users in different ways for certain social aims.  We believe the required effort for this will be significant, but it is likely to be useful for creating design patterns.

We believe the current state of our models will be helpful to both practitioners and researchers. As mentioned, the current models are capable of guiding system designs as well as research efforts.  The future work we have just described here enhance the utility of the models as well as maintain the models over time.

\section{Conclusion}
When it comes to social media design, which differences make a difference? CSCW theory has placed nearly all social media into a ``different time-different place'' cell of its traditional time-space matrix~\cite{johansen1988groupware,grudin1994computer}, and yet clearly social media express many meaningful similarities and differences. We introduce Form-From in this paper, offering a synthesized model derived from a larger 62-dimensional design space. Form-From offers a set of distinguishing features shared by designs in the same cell. We apply this model to demonstrate its analytical benefit, and invite the community to continue to extend and propose design space models of social computing systems going forward.

\begin{acks}
We would like to acknowledge Andrea Holber, Ricardo Jonathan, Simran Malhi, and Sumedha Kanumuri for their assistance with annotating platforms according to our 62-dimension model and helping to revise our codebook and definitions. We would also like to thank Steve Harrison, Jonathan Grudin, and Joseph Seering for their feedback and guidance on aspects of the literature review, model, and paper.

\end{acks}

\bibliographystyle{ACM-Reference-Format}
\bibliography{sample-base}

\appendix

\section{Full 62 Dimension Model}
\label{adx:model}

We list the entire 62-dimension model below grouped by the 10 high level categories.

\subsection{Category: Discussion Structuring}

This category describes how discussion data (the utterances that make up the discussion) is structured as whole into a data model, i.e., contained within data objects and the named relationships between data objects. In our distillation into the 2x3 Form-From model, this category is summarized into the dimension of ``Form''.

\subsubsection{Grouping of Utterances} This dimension describes how the utterances that make up a single discussion are grouped into data objects or containers within a system. Utterances can be grouped into multiple nested containers. 

Options (select all that apply): 
\begin{itemize}
\item Each user contribution forms a post, e.g., an email, a forum post, a comment, a message.
\item Stages/phases of a discussion, e.g., a system designed with explicit rounds of deliberation.
\item Grouping of posts into threads of conversation. Threads are groups of posts that form ``turns'' in a conversation, where one replies to the other; this doesn’t include quoting or linking to other posts.
\item Splitting long posts by a user into smaller connected chunks. Users on any platform with threads can do this by replying to themselves but on some platforms like Twitter, a thread of posts by the same author are incorporated into the system design and user experience.
\item Blocks of contiguous discussion time periods, such as meetings on Zoom. Usually for video/audio conversations that become a single data object storing the entire discussion time period.
\end{itemize}

\subsubsection{Multi-channeling}
Is all of the discussion focused into a single mode of interaction? Or are there parallel modes of interaction for the same discussion?

Options (Exclusive):
\begin{itemize}
\item  Single channel: only one channel for interaction, e.g., Messages (iOS) only has text chats.
\item Multiple channels: multiple parallel channels for the same discussion, e.g., Zoom meetings have a simultaneous video and chat channel, and Twitch and Instagram Live have a simultaneous video and chat channel.
\end{itemize}

\subsubsection{Threading/Partition of Replies}
\label{dim:reply-partition}
Many platforms offer a ``reply'' affordance and capture a relationship between a post and the post to which it replies.  If you record the exact post to which each post replies, the result is an infinitely deep tree.  But some platforms ignore the reply structure entirely, yielding a flat structure (often ordered chronologically). In an intermediate form, some systems capture reply links to a certain depth, but stop differentiating below that depth, yielding a fixed-depth tree.  For example, on Facebook today, a post has a first level of replies, but all posts descended from a particular reply are listed chronologically beneath that reply.

Options (exclusive): 
\begin{itemize}
\item  Fixed-depth reply tree, e.g., Slack (1 level of replies), Facebook (2 levels of replies).
\item  Infinitely deep reply tree, e.g., email, Reddit, Twitter.
\item  No replies, i.e., flat structure.
\end{itemize}

\subsubsection{Link Types}
Other than the common ``reply'' link for creating threads, which is covered in the above dimension, some platforms also provide structured mechanisms for other forms of linking, so that previous posts can be referenced as part of the current discussion.

Options (select all that apply): 
\begin{itemize}
\item  Citation: ability to cite another post within the discussion platform, e.g., 
 GitHub issues enable cross linking other posts, 4chan enable references to another post number, StackOverflow questions enable linking to another question.
\item  Embedded linking: ability for a post author to create their content in the context of another piece of content. This is commonly known as ``quote-posting'', such as Twitter quote-tweeting, Instagram embedded stories, and Tiktok duets and stitching.
\item  Pingbacks: displaying on a post a record of what other posts have cited that post; for instance, Blogger blogposts, Quora links when another question references a question, and Twitter links to all quote-tweets of a tweet.
\item  Argumentation links: a specific type of citation link that specifies the type of argument relationship between the two posts, i.e., ``refutes'' or ``supports''. This is often used in argumentation mapping systems such as the MIT Deliberatorium~\cite{Klein2012HowTH}. 
\item None.
\end{itemize}

\subsubsection{Post Types} 
\label{dim:post-types}
Are some posts distinguished from typical/standard posts?

Options (select all that apply): 
\begin{itemize}
\item Head post distinguished from replies: besides just being the first post in a thread, a system may incorporate additional designs to distinguish the first post, such as differences in appearance, format, post type, and discoverability compared to reply posts. Many of the below options go hand-in-hand with this option.
\item  Q\&A: some posts are questions, some are answers, e.g., Quora, StackOverflow.
\item  Argumentation System: posts have argumentation-related types, such as ``argument'' or ``evidence''.
\item  Pro-Con System: The initial post is the post to be debated, and subsequent posts are either of type ``pro'' or ``con'', e.g., on Kialo\footnote{\url{https://kialo.com/}} and ConsiderIt~\cite{Kriplean2012SupportingRP}.
\item  Summarizations of a post/thread: a special post type that is a summary of one or more other posts, e.g., Wikum~\cite{Zhang2017WikumBD}.
\item  Annotation: a specific post type that serves as an annotation to another post. One example is Twitter's Community Notes feature,\footnote{\url{https://twitter.com/i/communitynotes}} which annotates misinformation posts with corrections created by community members.
\item  None, i.e., only undiferentiated, generic posts or no concept of posts.
\end{itemize}

\subsubsection{Anchoring of Posts} 
\label{dim:anchoring}
What are posts anchored to? In other words, what are they posting on, if anything?

Options (select all that apply): 
\begin{itemize}
\item Anchoring post to text: posts are made on spans of text content. e.g., Google Docs, Medium, Nota Bene~\cite{Zyto2012SuccessfulCD}, and Hypothes.is have comments anchored to specific spans of document text or at a specific PDF location.
\item Anchoring posts to a particular point in time, e.g., SoundCloud has comments on song timestamps, YouTube has video comments anchored to a timestamp, most livestream apps like Twitch have chat messages anchored to specific timestamps.
\item Anchoring comment to space, e.g.,  virtual reality apps or physical spatial apps such as Foursquare anchor discussion to a specific virtual/physical space.
\item None.
\end{itemize}

\subsubsection{Enforced attachment of discussion to content item}
To what, if anything, are the discussion spaces as a whole attached? Here, we focus on enforced attachment, i.e., that conversation must be under some primary piece of content and cannot exist alone. We exclude the more generic case of discussions attached to a starting post that has a similar media format as discussion posts. This dimension also differs from the prior dimension in that attachment is to the content item as a whole.

Options (exclusive): 
\begin{itemize}
\item Attached to an article or document, e.g., the comments sections under many news articles.
\item Attached to a video, e.g., YouTube comments or TikTok comments.
\item Attached to an image, e.g., Instagram comments.
\item None.
\end{itemize}

\subsection{Delivery}

How are discussion items in the system delivered from user to user in the system and then presented?
In our distillation into the 2x3 Form-From model, this category is summarized into the dimension of ``From''.

\subsubsection{Main Location of Discussion} Where does discussion within the system reside and how does it arrive there? 

Options (Select all that apply):
\begin{itemize}
\item  Center: All discussion gets delivered to a single central globally addressable place, e.g., Twitter.
\item Rooms: Multiple distinct globally addressable places, e.g., Slack channels, forum boards.
\item Cliques: Multiple distinct places characterized only by membership and without a global address, e.g., direct messages, multi-party direct messages in Slack.
\item Mailboxes: Discussion gets delivered to personal inboxes without there being a shared central location to access all content, e.g., email, SMS.
\end{itemize}

\subsubsection{Ranking of Posts} 
\label{dim:ranking-of-posts}
Given the set of posts that can be received by a user, how are those top level ``posts'' ranked, ordered, or selected?

Options (Select all that apply):
\begin{itemize}
\item Chronological, e.g., Slack, WhatsApp.
\item Personalized algorithmic according to engagement, e.g., Facebook, Twitter.
\item Algorithmic according to collective voting, e.g., Reddit.
\item Editorial decision-making, e.g., NYTimes Editors' Picks.
\item Unclear, i.e., it is unknown or opaque in some way.
\item Other.
\item None.
\end{itemize}

\subsubsection{Ranking of Replies to a Post}
When a single ``post'' is followed by a set of comments associated with that item, how are the comments ordered?

Options (select all that apply): 
\begin{itemize}
\item Chronological, e.g., Slack replies.
\item Algorithmic according to collective voting, e.g., Reddit.
\item Unclear, i.e., it is unknown or opaque in some way.
\item By thread structure, e.g., Slashdot.
\item Other.
\item None, i.e., the system does not have comments underneath posts, instead everything is at one level, or does not have posts.
\end{itemize}

\subsubsection{Subscription}
\label{dim:subscription}
How can a user select which particular subset of posts they want to see from the body of content they are able to see? Selection could be a one-off event or something more like a subscription, where a user always wants to see content from that subset.

Options (select all that apply): 
\begin{itemize}
\item Posts from certain people, e.g., following people on Twitter, Facebook friends.
\item Specific topics/tags, e.g., following hashtags on Twitter.
\item All posts made to a group or channel, i.e., subscribe to a Facebook group, Discord channel.
\item Specific threads of conversation, i.e., subscribing to receive any new posts made to that thread.
\item Other.
\item None, i.e., subscription is not possible.
\end{itemize}

\subsubsection{Muting}
How can a user specify what or who they do not want to see from the body of content they are able to see? In other work, this has been characterized as personal moderation~\cite{Jhaver2023PersonalizingCM}.

Options (select all that apply):
\begin{itemize}
\item Posts from certain people, e.g., muting an account on Twitter even if following them. Note that blocking is covered under the Audience category, as it is concerned with both not seeing someone's content and also not allowing them to see your content.
\item Specific topics/tags/phrases, e.g., muting specific phrases on Twitter.
\item Replies to a post, e.g., the ability to mute a conversation that one was in on Twitter.
\item Score or degree related to some content metadata or characteristic. This involves the ability to set a threshold on a score of some kind, where posts below the threshold are hidden. Examples include post scores on an older version of Slashdot~\cite{Lampe2004SlashdotAB} or personal moderation settings for sensitive content on Instagram~\cite{Jhaver2023PersonalizingCM}.
\item Other.
\item None, i.e., users are not able to choose what they do not want to see.
\end{itemize}

\subsubsection{Findability} 
\label{dim:findability}
How can users find content they want to see if they have a particular kind of content in mind or if they want to go back to some specific content they've seen?

Options (select all that apply): 
\begin{itemize}
\item Text search, e.g., email clients, etc. If the platform is indexable, sometimes this search happens outside the platform, such as the common strategy of using Google Search to find Reddit threads.
\item User search, e.g., looking up a particular user in mind and then navigating to their page or their list of content.
\item Tags attached to content or hashtags embedded within textual content.
\item Navigating to specific boards, threads, and/or pages, e.g., 4chan, phpBB, Wordpress. Paginated forums typically have fixed size pages.
\item Scrolling back through one's feed, including with infinite scroll, e.g., Facebook Newsfeed.
\item Star/bookmarking of content for one's self, or personal saving of URLs to specific content.
\item Pinning, bookmarking, or saving of specific content for others, e.g., on Slack anyone can pin, while on Reddit mods can pin, and on Instagram, post authors can pin top replies.
\item None, i.e., there is no way to find particular content or go back to content. This can happen for instance in the case of systems with disappearing messages.
\end{itemize}

\subsubsection{Notifications of Content}
What controls the types of notifications that are received by the user in terms of notifications that point them to specific content? For this dimension, it's most helpful to only select the one that is most prevalent as we've found that many platforms have a mix of both.

Options (exclusive): 
\begin{itemize}
\item  User choice dominant.
\item Platform choice dominant, i.e., the platform pushes content into one's notifications that the user did not ask for.
\item No notifications.
\end{itemize}

\subsection{Temporality}

This category deals with all dimensions that relate to the timing and sychronicity of conversations.

\subsubsection{Requires Synchronous Interaction}
Does the discussion require that all participants be present live and interact synchronously? Many systems that are asynchronous can be used synchronously (e.g., instant messaging, Slack); here we are not interested in those, only ones that require the synchronous interaction.

Options (exclusive): 
\begin{itemize}
\item Requires synchronous, e.g., Zoom, Skype, FaceTime, OhYay.
\item Does not require.
\end{itemize}

\subsubsection{Discussion Endpoint}
Are there formal endings to a discussion? Or can it go on indefinitely?

Options (exclusive): 
\begin{itemize}
\item Automatic closure by time, e.g., We Are Dynamo~\cite{Salehi2015WeAD}, Precision Conference reviewer discussion period, timed Zoom meetings.
\item Automatic closure by activity level, e.g., 4chan threads close after inactivity.
\item Manually close the discussion, e.g., GitHub issues can be manually closed, Google Docs comment threads can be manually resolved, Zoom calls can be ended by the owner.
\item No ability to end a discussion, e.g., email~\cite{grandhi2016reply}, Twitter.
\end{itemize}

\subsubsection{Archiving}
What happens to the discussions afterwards? Are they saved and archived, or do they disappear? Some platforms offer both options, in which case: which is the default?

Options (exclusive): 
\begin{itemize}
\item Enforced ephemeral: content by default is always deleted (from the view of users). This includes 4chan, Snapchat, and Instagram Stories.
\item Enforced archived: content by default is always saved. Note: This combines platforms where you can delete comments manually and ones where you explicitly cannot, i.e., it is always recoverable. Examples include Wikipedia talk pages (cannot delete), Google Docs comments (cannot delete), Instagram comments (can delete).
\item Default archived: saved by default, but can be switched to ephemeral, e.g., Mailman lists.
\item Default ephemeral: ephemeral by default, but can be switched to archived, e.g., Zoom (can record the call), Slack (switching from the unpaid version to the paid version), Twitch (streams automatically get deleted but certain streamers can keep them around longer).
\end{itemize}

\subsubsection{Mutability} Can posts be edited, moved, or deleted by the creator of the post after creation?

Options (select all that apply): 
\begin{itemize}
\item Editable after posting, e.g., Slack and Facebook.
\item Moveable/retargetable after posting. 
\item Deletable after posting.
\item None.
\end{itemize}

\subsection{Audience}

This category is concerned with the flip-side of the category regarding Delivery, which is the audience of one's posts or discussions.
While the ``From'' dimension in the Form-From model comes most directly from the Delivery category and this category is more associated with an idea of ``To'', decisions regarding what is possible to do as a sender oftentimes mirror or constrain what is possible as a receiver, though they are not always the same. Thus, some of the concepts described within ``From'' have some overlap with dimensions in this category as well.

\subsubsection{Discussant Matching}
What is the process by which users get matched up with other users to have a discussion? In many platforms, this is via users, where users have control over who they want to talk to and can specify specific discussants or explore discussants in a shared space. The other option is some sort of algorithmic process for matching users or recommending users to each other. Many platforms have elements of both.

Options (select all that apply): 
\begin{itemize}
\item User-led exploration or selection.
\item Algorithmic matching of users or recommendation of users.
\end{itemize}

\subsubsection{Maximal Audience}
\label{dim:maximal-audience}
The upper bound on how many people can see a canonical post. That is, if every person spent infinite time looking at all the content they could see in the platform, how would we describe the set of content they can see?

Options (exclusive): 
\begin{itemize}
\item Bounded Group
\item Networked Neighborhood
\item Public within the Platform
\end{itemize}

\subsubsection{Delivery Guarantee/Typical Audience}
While a maximal audience could tell us theoretically how many people could see a comment if they wanted to, in practice, the audience of a comment is smaller. Who actually is made aware of the comment or guaranteed delivery?

Options (exclusive): 
\begin{itemize}
\item All members of group or recipient list, i.e., guaranteed delivery to all members of a group or recipient list. Examples include SMS, WhatsApp, Zoom, Slack, Twitter DMs, email, etc.
\item Unknown subset of group or recipient list, i.e., delivery is not guaranteed to all members of a group or recipient list. Examples include Facebook Groups, Instagram, TikTok, etc.
\end{itemize}

\subsubsection{Resharing Audience}
Who can the content be shared with, beyond the original audience?
This includes reshares that don’t modify the content (Re-post) as well as reshares that embed the original content in a new post (Quote-post).

Options (select all that apply): 
\begin{itemize}
\item Sender-specified people: can only be shared with a specific set of people, e.g., friends only.
\item Entire group: all members of the workspace can be forwarded the post but not further.
\item Public: everyone on the platform could eventually see the post.
\item None. This includes cases where reshares happen outside of the platform, such by sharing a URL via another platform.
\end{itemize}

\subsubsection{Limiting One's Audience}
What are the ways in which an author can limit the audience of their content?
On the flip side of a delivery guarantee, can an author guarantee (within the capabilities of a platform) that particular audiences do \textit{not} receive their content? This is commonly called blocking in most systems. 
Platforms can also allow have features to allow users to turn off resharing or replies to individual posts.

Options (select all that apply): 
\begin{itemize}
\item Authors can block other accounts from seeing their content.
\item Authors can turn off replies for one or more of their posts.
\item Authors can turn off resharing for one or more of their posts.
\item Other.
\item None.
\end{itemize}

\subsubsection{Technical Limit on Size}
What size group can technically participate in a conversation? Are there any technical limits on who can join or chime in?

Options (Select all that apply):
\begin{itemize}
\item Pairs (2), e.g., ChatRoulette, phone calls.
\item Small to medium group (10s), e.g., iMessage has a limit of 32 people.
\item Large group (100s), e.g., Zoom unpaid has a limit of 100 people.
\item Large community (1000s), e.g., WhatsApp Groups has a limit of 1,024 people and WhatsApp Community enables up to around 21,000 members.
\item Internet scale, i.e., no limit theoretically.
\end{itemize}

\subsubsection{Targeting of Posts}
Beyond simply sending posts to specific people or to specific channels, how can post authors target their posts towards someone or someplace and how is that targeting made apparent in the discussion platform?

Options (select all that apply): 
\begin{itemize}
\item Target to specific people/audience, including @-mentioning specific people or having an explicit ``to''.
\item Differentiated targeting to different audiences; for instance, in email, the feature of separating out To, CC, and BCC~\cite{Zhang2020ConfiguringAA}.
\item Target to specific topics, including hashtags.
\item None.
\end{itemize}

\subsubsection{Audience Solicitation Mechanisms}
In what ways, if any, does the platform provide for solicitation of certain audiences?

Options (select all that apply): 
\begin{itemize}
\item Solicitation of replies or responses. Some Q\&A sites such as Quora solicit responses from certain people regarding questions that are still open and suited for them to answer.
\item Inserting out-of-band content into view. Many ads on social platforms operate in this way by being inserted into ongoing conversation to reach the audience level and demographic that was paid for.
\item Tagging users. Authors as well as audiences may solicit users by tagging them.
\item None.
\end{itemize}

\subsubsection{Backchanneling}
Are secondary conversations allowed during the main conversation?

Options (exclusive):
\begin{itemize}
\item
Enable backchanneling (e.g., private message as part of a larger conversation). For instance, this includes the ``Whisper'' feature in OhYay or private direct message in Zoom chat. Indeed, almost all apps have some kind of direct message capability that allows one to take a more public conversation to a more private one. However, in these cases, the feature is more closely integrated with the main conversation, whereas direct messaging is more like a distinct product surface on other apps.
\item None.
\end{itemize}

\subsubsection{Spatial Proximity}
Does the discussion start or depend on the location of a user? This can include achieving a certain user distance in order for the discussion to be possible to perceive at all.

Options (exclusive): 
\begin{itemize}
\item Spatial distance to discussion affects strength of audio, size of content. Most VR, AR, and apps involving physical location (e.g., Pokemon Go) have this characteristic.
\item None.
\end{itemize}

\subsection{Membership}

What defines who can be a member in the discussion? What kind of memberships are available, and what do they enable?

\subsubsection{Roles/Differentiation in Capabilities}
What are the allowable social roles \textit{supported} by the system?  There may be informal roles used by participants using the system, but the roles to be coded have direct support by system features.  There must be some role in the system (likely at least two), so None is not an option.

Options  (select all that apply): 
\begin{itemize}
\item Regular users:  the most generic role in terms of privileges.
\item Moderators:  this role has the privileges of regular users but also can handle content changes.
\item Admins:  this role has regular user and moderator privileges but also controls the system, users, and other system capabilities. 
\item Other:  there are additional roles. For instance, there may be gradations in the handling of content for moderators (e.g., cannot modify, only delete, content; can delete only some content). There may also be  multiple levels of any of these roles (e.g., admins that can handle user complaints and admins who have code access). As another example, Matrix allows every user to be placed on a 1--100 level power ranking.
\end{itemize}

\subsubsection{Traceability/Account Index}
How is one's real identity known (if it is known) to admins? A phone number, street address, specific email address, some organizational affiliation (e.g., a university), one's name, a credit card number, IP address, or some combination can be tied to a user account.  In some locales, the user account can also be tied to a national ID.  
Note that these are not exclusive; a system can have more than one (although presumably None and one of the others is not allowed in the system).

Options (select all that apply): 
\begin{itemize}
\item
Tied to phone number
\item Tied to national ID
\item Tied to street address
\item Tied to specific email:  a specific email address is tied to the user’s account
\item Tied to university or other organizational affiliation
\item Tied to name:  a full name is tied to the user’s account
\item Tied to credit card
\item Tied to IP address
\item Generic verified identity (process is opaque or multi-faceted)
\item Other.
\item None (should be used only in cases where users have complete anonymity).
\end{itemize}

\subsubsection{Criteria for Joining}
By what criteria are users able to join a site constructed in the system?  Admins or others may need to decide membership in the site (i.e., not everyone can join).  If so, there will be some criteria upon which the decision to allow a specific user to join will be made.

Options (select all that apply): 
\begin{itemize}
\item
Geofencing:   the prospective member must live/enhabit some physical bounded location.
\item Members must vouch for you:  at least one current user must approve the prospective member.
\item Donation:  some payment  (e.g., \$5/month) allows a prospective member to join.
\item Personal characteristics (e.g., demographics):  some personal characteristic (e.g., age,  occupational group, educational background) allows a prospective member to join.
\item Organizational membership:  a prospective member must be an employee or some other formal affiliation with an organization, company, or university.
\item Company customer (e.g., customer support forum).
\item Other:  if there is another criteria by which people are judged to be suitable site members, provide the criteria.
\item None:  anyone can join.
\end{itemize}

\subsubsection{Relationship between Person and Account}
You guessed it:  There are people and there are accounts on the site.  There should be system support for both.  What’s the relationship between the two? Note that none is not an option. Also note that we are explicitly interested in system support for the relationship, not users repurposing a system designed in a particular way, i.e., by sharing passwords.

Options (select all that apply): 
\begin{itemize}
\item One Account One Person
\item 
Multiple Accounts per Person
\item 
Multiple People per Account
\end{itemize}

\subsubsection{Criteria for Staying}
You guessed it: People want to stay around the site.  What do they need to do to be allowed to stick around?

Options (select all that apply): 
\begin{itemize}
\item 
Fixed duration membership according to time.
\item 
Must meet activity threshold:  examples include log in a certain number of times, have enough points.
\item
Must retain affiliation (continue to satisfy criteria for joining).
\item 
Must continue as customer.
\item 
Other. 
\item 
None:  A user can stick around for as long as they want or until an admin/owner/platform kicks them out.
\end{itemize}

\subsection{Moderation}

How content is moderated and curated on platforms has significant impact on the experience for users. 
We define moderation to be something someone does that affects what other people see, not just what they themselves see, which could be seen just as personal filters or controls, which are covered in other categories.
Here we also focus on the technical supports for moderation that different systems provide, while not considering specific practices that different moderation or Trust \& Safety teams employ, which can change policy regularly.
We also use a broad definition of moderation that includes content curation actions as defined in Grimmelman~\cite{Grimmelmann2015TheVO}.

\subsubsection{Moderation Model}
How is moderation structured on the platform? None is not an option, as platform moderation would be the default.

Options (select all that apply): 
\begin{itemize}
\item 
Networked/user-led moderation: A user or organization can moderate/create moderation settings for another person. Examples: SquadBox~\cite{Mahar2018SquadboxAT}, shared blocklists that were once built for Twitter~\cite{Jhaver2018OnlineHA}, and BlueSky's composable moderation and custom feeds.\footnote{\url{https://blueskyweb.xyz/blog/4-13-2023-moderation}}
\item Community moderation: members of the group are empowered to moderate their own group.
Examples: Subreddits, Discord servers, Mastodon instances.
\item 
Platform moderation: the platform, in a centralized way via human annotation, algorithms, or a combination, moderates content. Examples: YouTube, Facebook, Twitter, Threads.
\end{itemize}

\subsubsection{Moderation Decider (Who)}
Who is empowered to engage in moderation on the platform? None is not an option, as the creators of the platform must always make basic moderation decisions.

Options (select all that apply): 
\begin{itemize}
\item 
Members: regular users can engage in moderating for their friends, followers, posters on their content, or their community. Examples include being a community moderator of Discord or moderating the comments on one's Youtube video or Instagram image post.
\item Contract human moderators: professional labor force tasked with carrying out existing rules but typically not innovating on rules.
\item Platform creators/employees: top-down decisions from the platform's Trust \& Safety team, developers, or leadership.
\item Third-party organization: independent organization who provides recommendations to the platform. Examples include Facebook's Oversight Board or fact-checking organizations that contract with social media companies. In some cases, the government provides recommendations as well.
\item Algorithms: can be simple or complex algorithms that leverage votes and other signals as well as perform inference to conduct moderation.
\end{itemize}

\subsubsection{Actions Available to User Moderators (What)}
What specific powers do moderators have? We focus here on moderators who are users of the platform and not employed by the platform or a third-party organization tasked with doing this as part of their work, as employees, contractors, and third-party organizations typically use custom-built internal tools to do their work. Those tools would fall outside of our scope of examining the design space of social media systems.

Options (select all that apply): 
\begin{itemize}
\item 
Create privileged content (e.g., content highlighted as coming from a moderator)
\item
Modify content (e.g., change others' posts)
\item
Delete content
\item
Annotate content
\item
Invite members to join
\item
Ban, suspend, or shadowban members
\item
Promote, demote content (e.g., star, pin, resolve, endorse)
\item
Remove members' abilities to perform specific actions
\item
Other.
\item
None (there are no community moderators).
\end{itemize}

\subsubsection{Moderation Raised (How Triggered?)}
What processes are followed to determine whether to engage in a moderation action?

Options  (select all that apply): 
\begin{itemize}
\item 
Manual by moderator, e.g., Reddit moderators can choose to moderate individual messages.
\item 
Reported by users, e.g., a user flags a post on Facebook that get surfaced to moderators.
\item 
Human-written rules: user or moderator written rules such as if/then statements, regular expressions, word lists that can be checked automatically. Example: Reddit AutoModerator.
\item 
AI-learned rules: rules learned from labeled data that can be executed automatically. Example: Facebook's COVID-19 misinformation detection model.
\end{itemize}

\subsubsection{Moderation Timing (When?)}
When are moderation actions taken?

Options (exclusive): 
\begin{itemize}
\item 
Proactive: moderation occurs before content goes live. Example: Mailman email list moderation requires an admin to approve any post before it goes live; YouTube algorithms remove videos with hate speech before they are posted live.
\item
Reactive: moderation occurs after content goes live or gets reported.
\end{itemize}

\subsubsection{Transparency of Moderation}
Who can see the moderation rules and how they are carried out?

Options (select all that apply): 
\begin{itemize}
\item 
Rules visible: the rules used for moderation decisions are publicly visible
\item
Implementation visible: the procedures/discussions/algorithms used to execute the rules are publicly visible. Example: Wikipedia discussions happen in public Talk pages.
\item
Opaque: neither rules nor their implementation are visible.
\end{itemize}

\subsection{Governance}

Unlike Moderation, which is all about taking actions over content or users and questions of who, what, and when regarding those actions, Governance is more broadly about power and decision-making within social media and system designs that support governance. In this set of dimensions, we focus on high level governance questions that characterize the system as a whole but other work has delved into more detailed dimensions of governance, such as the partitioning and interplay of power between different governing entities within a platform in cases where it is decentralized~\cite{jhaver2023decentralizing}.

\subsubsection{Access to Power}
How do users on the platform get more capabilities for control over the activity on the site?  
It does not matter, for this design dimension, whether the people are volunteers or paid, but we are not including people appointed in salaried positions by the platform (developers, product managers, etc.), only community members/users of the platform.

Options (select all that apply): 
\begin{itemize}
\item 
Reputation System:  There exists some algorithm that decides to promote users to additional capabilities, and that algorithm uses some ranking system for reputation.
\item
Election Process:  Moderators and other roles can be elected by those with the same capabilities or those with fewer. An example is English Wikipedia, which holds elections for many of their roles~\cite{Leskovec2010GovernanceIS}.
\item
Promotion/Appointment:  users may be promoted into moderator or admin roles, or in the general case, people may be promoted through hierarchical ranks by those above them in the hierarchy.  A typical process is that users that are active get promoted to moderators, and good moderators may get promoted to admins. This option is common in space-based social media today~\cite{Schneider2021AdminsMA}.
\item None:  no access to power for regular users. This option is common in network- and commons-based social media.
\end{itemize}

\subsubsection{Conflict Resolution}
How are conflicts on the site between users or communities mediated and resolved?  

Options (select all that apply): 
\begin{itemize}
\item Mediation:  There is some known process where both parties in the conflict come together and try to work out their differences, perhaps with someone appointed as a conflict mediator.
\item
Voting:  The users on the system vote on the issue.
\item
Judicial process:  There are a group of users (perhaps special administrators or a jury of users) who decide on discipline issues based on a set of rules or policies.
\item
Appeal to authority:  There are administrators or leaders of some kind who decide discipline issues.  These decisions may or may not be explained.
\item
None (unlikely but possible).
\end{itemize}

\subsubsection{Decision-makers}
Who makes decisions on the platform regarding platform-wide policy?  

Options (select all that apply): 
\begin{itemize}
\item
Users  (democracy). For example, imagine a direct vote or ``citizen'' assembly to decide on a possible policy change.
\item 
A subset of elite users  (oligarchy):  Some group, perhaps based on roles like administrators and moderators or perhaps based on other attributes such as reputation points, decide.
\item
Platform operators  (autocracy):  The owners of the platform decide.
\item
None (unlikely but possible).
\end{itemize}

\subsection{Extensibility}

A social media platform can be transformed by opening itself up for extension by third-party tools, customization by users, and interoperability with other platforms.

\subsubsection{Bots}
Can developers and/or users utilize bots that are supported by the platform?

Options (select all that apply): 
\begin{itemize}
\item
Conversational Bots: Bots start conversations (e.g., chatbots to assist users).
\item
Moderation Bots: Bots moderate content (e.g., identify banned content).
\item
Other.
\item
None or rarely used.
\end{itemize}

\subsubsection{APIs}
What APIs are available through the platform?

Options (select all that apply): 
\begin{itemize}
\item
API Access to Extract Data: APIs can be used to read data.
\item
Bot API: Developers can create bots to listen for and post content.
\item
API Access to CRUD Discussion: APIs can be used to access or modify data (CRUD operations: create, read, update and delete). This would enable users to use custom clients.
\item 
Federation API: APIs can be used to share content and follow people across different instances of a social media system using the same protocol.
\item
None.
\end{itemize}

\subsubsection{Platform Customization}
What options are available to users and communities for customization?

Options (select all that apply): 
\begin{itemize}
\item
User front-end customization: a user can change the look and feel of the platform for themselves via personal ``skins''. The most common example of this is light vs dark mode on many apps.
\item 
User client choice:  users can select a particular client, e.g., the many Mastodon clients that can be used to access Mastodon instances.
\item
Community front-end customization: communities can change the look and feel of some aspect of the platform, e.g., community-specific flairs on Reddit. Another example is pre-set widgets that communities can decide to turn on.
\item
Community server choice: a community can select a particular server implementation to have the features they want, e.g., different forks of the Mastodon server code.
\item
Custom moderation or curation: a user can choose to customize some element of moderation (e.g., via personal controls for unwanted content~\cite{Jhaver2023PersonalizingCM}) or curation (e.g., via selecting custom feed sorting algorithms on BlueSky).
\item
Other.
\item None.
\end{itemize}

\subsubsection{URL linking}
What can be accessed via URL links outside of the platform?

Options (select all that apply): 
\begin{itemize}
\item
URL link to posts (that contain replies), threads, channels, or other groups of comments.
\item
URL link to every comment.
\item
None.
\end{itemize}

\subsection{User/Group Signaling}

Social media platforms incorporate and display many signals at the user and group level to convey useful contextual information to users.

\subsubsection{User Identification and Verification Markers}
What options are available for identifying a user on the platform?

Options (select all that apply):
\begin{itemize}
\item
Animated avatar, e.g., in VR.
\item Profile image.
\item Display name.
\item Username.
\item Badge, e.g., a ``verified'' badge or ``blue check'' that someone is who they say they are.
\item Organizational affiliation or occupation, e.g., verified labels that someone is a politician or state-sponsored news organization.
\item Other.
\item None.
\end{itemize}

\subsubsection{Anonymity to Other Users}
How known are regular users to other users?  (This is not about being known to platform employees or server admins, only to other users.)  Anonymity means a message from that user cannot be tied to *any* account. Pseudonymity means one cannot tie an account to any specific person.  

Options (select all that apply): 
\begin{itemize}
\item Anonymity enforced: the system does not attach messages to any identifier (except via self-disclosure), e.g., 4chan.
\item Pseudonymity enforced: the system attaches messages to an identifier but the identifier is assigned by the system. This way, the system does not allow a user's identity to be known (except via self-disclosure). For example, Twitter Community Notes members have a codename.
\item Self-selected identifier:  users can decide for themselves whether their identity is anonymous, pseudonymous, or known. 
Most platforms are here.
\item Temporary or post-level anonymity or pseudonymity: the user can decide for themselves whether to cloak their identity in a particular time period or for a particular post. For example, classroom forums like EdStem and Piazza allow instructors to enable posts to be anonymous if the user wants to submit an anonymous question.
\item Identity cloaked to public but revealed to receiver/other specific users. For instance, EdStem and Piazza instructors can see the identity of anonymous posts.
\item Identity enforced: no anonymity or pseudonymity permitted. Examples include Facebook with their Real Name Policy and NextDoor.
\end{itemize}

\subsubsection{Differentiated Roles/Faces/Profiles revealed to others}
Are people allowed to differentiate the roles, level of anonymity, and profile information they present/reveal to different other users? Or do users look the same to every other user?

Options (exclusive):
\begin{itemize}
\item Yes
\item No
\end{itemize}

\subsubsection{Activity Indicators (What?)}
What indicators does the platform provide to help you find out what kinds of activity are going on by other people on the platform at that current moment?

Options (select all that apply): 
\begin{itemize}
\item 
Individual presence, e.g., “I am busy/away”, “I am here right now”.
\item
Individual user activity, e.g., “Person A is typing”.
\item 
Individual location marker, e.g., live cursor on Google Docs and Figma.
\item 
Live avatar presence in channel or space, e.g., Discord, Gather.town.
\item 
Collective activity, e.g., “5 people are typing”, group radar and marbles in Loops/Babble prototype.
\item
Collective location marker.
\item Other.
\item None.
\end{itemize}

\subsubsection{Activity Indicators (How?)}
What are the different ways that a user finds out about other users’ activity?

Options (select all that apply): 
\begin{itemize}
\item 
Out of band notifications about activity, e.g., ResearchGate emails.
\item 
Integration of activity information into a feed, e.g., Facebook feed incorporates activity information.
\item  Log of activity, e.g., Slack or Discord sends a message when someone has left the chat, Wikipedia talk pages record all activity, Tumblr captures logs of activity on a post.
\item 
Badge or tag or icon next to an account, e.g., away status in AIM.
\item 
Temporary pop-up indicator that appears near where the activity is occurring, e.g., ``So-and-so is typing''.
\item Other.
\item 
None.
\end{itemize}

\subsubsection{Content Signals}
How can users express their opinions of the content?

Options (select all that apply):
\begin{itemize}
\item 
Vote/Reaction/Label
\item 
Personal Endorsement
\item Other
\item 
None
\end{itemize}

\subsubsection{Organizational Meta-Data}
What meta-data is available that conveys content categorization/organization? 

Options (select all that apply): 
\begin{itemize}
\item 
Optional label/tag/highlight on post
\item 
Closed vocabulary required categorization
\item 
Open vocabulary required categorization
\item
None
\end{itemize}

\subsubsection{Post Verification}
How is a post verified?

Options (select all that apply):
\begin{itemize}
\item Needs to be in a certain location (Foursquare check-in) or checked in to a place (Yelp review).
\item Needs to have purchased the thing (Amazon review).
\item Other.
\item None.
\end{itemize}

\subsubsection{Expression related to identity/account}
Is there an expression that a user can tie to their account beyond their identity marker?

Options (select all that apply): 
\begin{itemize}
\item
Status message.
\item 
Personal bio.
\item 
Signature.
\item Other.
\item 
None.
\end{itemize}

\subsubsection{Community self-expression}
Is there an form of expression that a group of users can tie to their community beyond their name/address for the community?

Options (select all that apply): 
\begin{itemize}
\item
Group description.
\item
Group pinned or bookmarked content.
\item
Code of Conduct.
\item
FAQ.
\item Logo.
\item Other.
\item None.
\end{itemize}

\subsection{Content}

Finally, we turn to content. This category involves the ways that social media systems vary according to the particulars of the content itself and how it is authored.

\subsubsection{Communication Media/Content Type}
What are the different mediums and artistic elements used by the creator to express themselves on platform?

Options (select all that apply): 
\begin{itemize}
\item
Text.
\item
Audio.
\item
Video.
\item
Movements of a digital avatar, e.g., in VR.
\item
Screensharing.
\item
Drawing/whiteboard.
\item
Images.
\item
Quoting and linking to other posts.
\item
Styling formatting of text + inline media (HTML, CSS).
\item
Glyphs.
\item
Emojis.
\item Other.
\end{itemize}

\subsubsection{Authoring Creative Constraints}
Limits that the platform places on the user when creating and designing posts. 

Options (select all that apply): 
\begin{itemize}
\item
Character count limit.
\item
Time limit on duration of post.
\item
Image size limit.
\item
Media playback duration limit.
\item
Other.
\item
None.
\end{itemize}

\subsubsection{Authoring Support}
The methods to which the platform provides resources to creators to convey the intended message and design the post.

Options (select all that apply): 
\begin{itemize}
\item AI-mediated communication: AI suggestions for what to write, e.g., Gmail SmartReply.
\item Filter/ Lenses, e.g., any changes to color, shapes within the post/image.
\item Overlays: any items that can be placed on top of the post (e.g., stickers, drawings).
\item Accessibility tools:  opportunities provided to make the post more accessible, e.g., by adding alt text.
\item Remix: use content provided by the platform to add into the post you are making, e.g., adding a song or audio to your Instagram Story or Tiktok video.
\item Other.
\item None.
\end{itemize}

\subsubsection{Structuring of a Post}
How the post is structured to effectively communicate the idea.

Options (select all that apply): 
\begin{itemize}
\item 
Title/summary followed by the main body.
\item Multi-part Post.
\item Other.
\item None.
\end{itemize}

\subsubsection{Post Meta-Data}
What meta-data does the system allow for a post (including both system-added and author-added)?

Options (select all that apply): 
\begin{itemize}
\item 
Author.
\item 
Timestamp.
\item
Physical location.
\item Other.
\item None.
\end{itemize}

\subsubsection{Drafting}
Can posts and the different content within the post be drafted and posted at a later time.

Options (exclusive):
\begin{itemize}
\item 
Storing and retrieving drafts of messages.
\item 
None.
\end{itemize}

\section{Coding of Historical Systems}\label{apx:history}

This list is coded from the historical list of social media platforms on Wikipedia. The version of the list that we coded is permanently available at \url{https://en.wikipedia.org/w/index.php?title=Timeline_of_social_media&oldid=1165195616}. ARPANET and Co-star were excluded for not meeting our paper's definition of a social media platform.

\begin{longtable}[htbp]{llll}
\hline
\textbf{System} & \textbf{Year} & \textbf{Form} & \textbf{From} \\ \hline
\endfirsthead
\hline
\textbf{System} & \textbf{Year} & \textbf{Form} & \textbf{From} \\ \hline
\endhead
Talkomatic & 1973 & Flat & Spaces \\
TERM-Talk & 1973 & Flat & Spaces \\
ARPANET & 1974 & -- & --  \\
PLATO Notes & 1974 & Threads & Spaces \\
BBS & 1980 & Flat & Spaces \\
FidoNet & 1984 & Flat & Spaces \\
IRC & 1988 & Flat & Spaces \\
Classmates.com & 1995 & Flat & Spaces \\
Bolt.com boards & 1996 & Threads & Spaces \\
Bolt.com chat & 1996 & Flat & Spaces \\
LunarStorm & 1996 & Flat & Networks \\
ICQ & 1996 & Flat & Spaces \\
SixDegrees.com & 1997 & Flat & Networks \\
AOL Instant Messenger & 1997 & Flat & Spaces \\
Open Diary & 1998 & Threads & Spaces \\
LiveJournal & 1999 & Threads & Networks \\
Yahoo! Messenger & 1999 & Flat & Spaces \\
MSN Messenger & 1999 & Flat & Spaces \\
Habbo Hotel & 2000 & Flat & Spaces \\
Friends Reunited & 2000 & Flat & Networks \\
Windows Messenger & 2001 & Flat & Spaces \\
Friendster & 2002 & Flat & Networks \\
4chan & 2003 & Threads & Spaces \\
LinkedIn & 2003 & Flat & Networks \\
Hi5 & 2003 & Flat & Networks \\
Xing & 2003 & Flat & Networks \\
MySpace & 2003 & Flat & Networks \\
Skype & 2003 & Flat & Spaces \\
Flickr & 2004 & Threads & Networks \\
Tagged & 2004 & Threads & Networks \\
Facebook & 2004 & Flat & Networks \\
Orkut & 2004 & Flat & Networks \\
Reddit & 2005 & Threads & Spaces \\
YouTube & 2005 & Threads & Networks \\
Qzone & 2005 & Threads & Networks \\
Renren & 2005 & Threads & Networks \\
Bebo & 2005 & Flat & Networks \\
Twitter & 2006 & Threads & Networks \\
VK & 2006 & Threads & Networks \\
NK.pl & 2006 & Threads & Networks \\
Tumblr & 2007 & Threads & Networks \\
FriendFeed & 2007 & Threads & Networks \\
Justin.tv & 2007 & Flat & Spaces \\
Sina Weibo & 2009 & Threads & Networks \\
Quora & 2010 & Threads & Spaces \\
Instagram & 2010 & Threads & Networks \\
Path & 2010 & Threads & Networks \\
Pinterest & 2010 & Flat & Commons \\
Snapchat & 2011 & Threads & Networks \\
Google+ & 2011 & Threads & Networks \\
Keek & 2011 & Threads & Networks \\
Twitch & 2011 & Flat & Spaces \\
Tinder & 2012 & Flat & Commons \\
YikYak & 2013 & Threads & Spaces \\
8chan & 2013 & Threads & Spaces \\
Vine & 2013 & Threads & Networks \\
Patreon & 2013 & Threads & Networks \\
Slack & 2013 & Flat & Spaces \\
Google Hangouts & 2013 & Flat & Spaces \\
Musical.ly & 2014 & Threads & Networks \\
Periscope & 2015 & Threads & Networks \\
Beme & 2015 & Threads & Networks \\
Meerkat & 2015 & Threads & Networks \\
Discord & 2015 & Flat & Spaces \\
Triller & 2016 & Threads & Networks \\
Mastodon & 2016 & Threads & Networks \\
Gab & 2016 & Threads & Networks \\
Co-star & 2017 & -- & --  \\
Pillowfort & 2017 & Threads & Networks \\
TikTok & 2017 & Threads & Commons \\
Parler & 2018 & Threads & Networks \\
BeReal & 2020 & Threads & Networks \\
Clubhouse & 2020 & Flat & Spaces \\
Gettr & 2021 & Threads & Networks \\
Truth Social & 2021 & Threads & Networks \\ 
Nostr & 2023 & Threads & Networks \\ 
Threads & 2023 & Threads & Commons \\ \hline
\end{longtable}

\end{document}